\documentclass[aps,twocolumn,secnumarabic,nobalancelastpage,amsmath,amssymb,
longbibliography]{revtex4-1}
\usepackage{amsmath}
\usepackage{graphics}      
\usepackage{graphicx} 
\usepackage{subcaption}   
\usepackage{url}           
\usepackage{bm}            
\usepackage[english]{babel}
\usepackage{pgfplots}
\usepackage[utf8]{inputenc}

\usepackage{ragged2e}
\DeclareUnicodeCharacter{2212}{-}
\begin{document}
\title{Charge Constraints of Macroscopic Dark Matter}
\author         {Jagjit Singh Sidhu}
\affiliation    {Physics Department/CERCA/ISO Case Western Reserve University
Cleveland, Ohio 44106-7079, USA}
\date{\today}

\begin{abstract}
Macroscopic dark matter (macros) refers to a broad class of alternative candidates
to particle dark matter with still unprobed regions of parameter space.
Prior work on macros
has considered elastic scattering
to be the dominant energy transfer mechanism in deriving constraints
on the abundance of macros for some
range of masses $M_x$ and (geometric) cross-sections $\sigma_x$
However,
macros with a
significant amount of electric charge
would, through
Coulomb interactions,
interact 
strongly enough
to have produced
observable signals
on terrestrial, galactic
and cosmological scales.
We determine the expected {   phenomenological} signals
and 
constrain the corresponding regions 
of parameter space,
based on the lack of these
signals in observations.
\end{abstract}
\maketitle

\section{Introduction}
The is strong evidence that
dark matter is the dominant
form of matter in the Universe
(see e.g. \cite{PDG}).
Dark matter 
explains several phenomena on 
both galactic
and cosmological scales \cite{PDG},
from the shape of 
galaxy rotation curves
to the history of 
structure formation. 
However, the precise nature of 
dark matter
remains one of the big unsolved problems
in cosmology.

New fundamental particles, not included in the Standard Model of particle physics, 
are popular candidates because they often arise in models of
Beyond the Standard  Model physics 
that were 
invented for independent reasons 
(e.g. the axion \cite{axion,Wilczek_axion,Weinberg_axion}).
However, it remains an open possibility that dark matter is comprised instead 
entirely of macroscopic bound states. 

Such bound states would avoid strong constraints on the self-interactions of dark matter by virtue of their low number density instead of any intrinsic weakness in their non-gravitational couplings. 
One such open possibility is that dark matter is comprised of macroscopic bound states of quarks or hadrons, as first proposed by Witten \cite{PhysRevD.30.272} 
as products of a first-order QCD phase transition, and later Lynn, Nelson, and Tetradis \cite{LYNN1990186} and Lynn \cite{1005.2124} again,
who argued in the context of SU(3) chiral perturbation theory that ``a bound
state of baryons with a well-defined surface may conceivably form in the presence of kaon condensation.'' This would place the dark matter squarely within the Standard Model.
Others have suggested non-Standard Model versions of such objects and their formation, for example incorporating the axion \cite{hep-ph/0202161}.
{  Additionally, it has been noted in reference \cite{Pontn2019} that in a
simple Higgs-portal complex scalar dark matter model, a non-topological soliton
state exists for dark matter. This work also considered one possible
mechanism to produce macroscopic dark matter soliton states from early-universe dynamics, i.e. a
first-order phase transition of electroweak (EW) symmetry \cite{Pontn2019}}

Due to their large mass and low number density, macro detectors must be extremely large,  experience extremely long integration times or
be proficient at accumulating 
dark matter due to e.g.
gravitationally-enhanced Sommerfield enhancement as in
white dwarfs and neutron stars,
to overcome the macros' low flux compared to typical particle dark matter. 

In recent years the author and
collaborators have determined
the regions of macro
parameter space that cannot 
constitute all of the
dark matter based on several
null observations
in various experiments
\cite{jacobs2015macro,jacobs2015resonant, 1907.06674, 1908.00557}.
We have also discussed
further ways to probe 
more of the remaining parameter
space \cite{Sidhu:2018auv,1905.10025}.
These works assumed the dominant
interaction to be
elastic scattering
and the interaction cross-section,
was taken
to be the geometric cross-section of the macro, i.e.
$\sigma_{elastic} = \sigma_x$.
For more details on recent
work involving macros
as viable dark matter candidates,
we refer the reader to
the works cited above and
references therein.
{  
However, we begin
by first briefly reviewing the 
existing constraints on
derived
from previous work.

For macro
masses $M_x \leq 55\,$g careful examination of specimens of
old mica for tracks made by passing dark matter \cite{DeRujula:1984axn,Price:1988ge}
has ruled out such objects as the primary dark-matter
candidate (see Figure 1). 
For even smaller masses $M_x \leq 55 \times 10^{-4}\,$g, a similar constraint was obtained \cite{1805.07381} using the 
MACRO detector \cite{hepex0207020}
For $M_x \gtrapprox 10^{21}\,$g, a variety of microlensing searches have constrained the abundance of
macros \cite{Alcock2001,astro-ph/0607207,0912.5297,Griest2013,Niikura2019} from a lack
of magnification of sources
by a passing macro along the line
of sight of the observer. {\it{   The most recent
lensing constraints from M31 have recently been corrected taking
into account a more realistic model for the source stars in M31 \cite{1910.01285}.}}

A large region of parameter space was constrained by considering thermonuclear runaways triggered by macros incident on white dwarfs 
\cite{1805.07381}. Dark matter-photon elastic interactions 
were used together with Planck cosmic
microwave background data to constrain macros of sufficiently high reduced
cross-section $\sigma_x/M_x$ \cite{1309.7588}.
Prior
work had already constrained a similar range of parameter space by showing that 
the consequence of dark matter
interactions with standard model particles is to dampen
the primordial matter 
fluctuations and essentially erase
all structures below a given scale (see e.g. \cite{Bhm}).
The region of parameter
space where macros would have produced a devastating
injury similar to a gunshot wound on the carefully monitored population of the western world was also recently
constrained \cite{1907.06674}.

Recently, together with collaborators, we suggested
how ultra-high-energy cosmic ray detectors that exploit
atmospheric fluoresence could potentially be modified to
probe parts of macro parameter space \cite{Sidhu:2018auv}, including
macros of nuclear density. This analysis
has led to constraints already being placed
using networks of cameras that were originally built
to study bolides, i.e. extremely bright meteorites with
absolute magnitude
\cite{1908.00557}. We have also suggested
how the approach applied
to mica \cite{DeRujula:1984axn,Price:1988ge} could be applied to a larger, widely available sample of appropriate rock \cite{1905.10025}, and used to search for
larger-mass macros. 
We have also identified additional
regions of parameter space
constrained by the duration
between back-to-back
superbursts (thermonuclear
runaway on the outer surface of
a neutron star) \cite{1912.04053}.

It is unlikely that macro masses beyond $\sim 10^{9}\,$g
could be probed by any purpose-built terrestrial detector assuming even an observation time of a century and
a target area the size of the Earth. Terrestrial probes
(eg. ancient rocks \cite{DeRujula:1984axn,Price:1988ge,1905.10025}) could have been continuously exposed for up to $\sim 3 \times 10^9$ years, but we are unlikely
to carefully examine the more than $1\,$km$^2$
that would be
needed to push beyond $M_x \sim 10^9\,$g. It will therefore require innovative thinking about astrophysical probes (eg.
\cite{1805.07381}) to probe the
remaining unprobed parameter space
at the very highest masses.}

 \begin{figure*}
  \includegraphics[width=\textwidth]{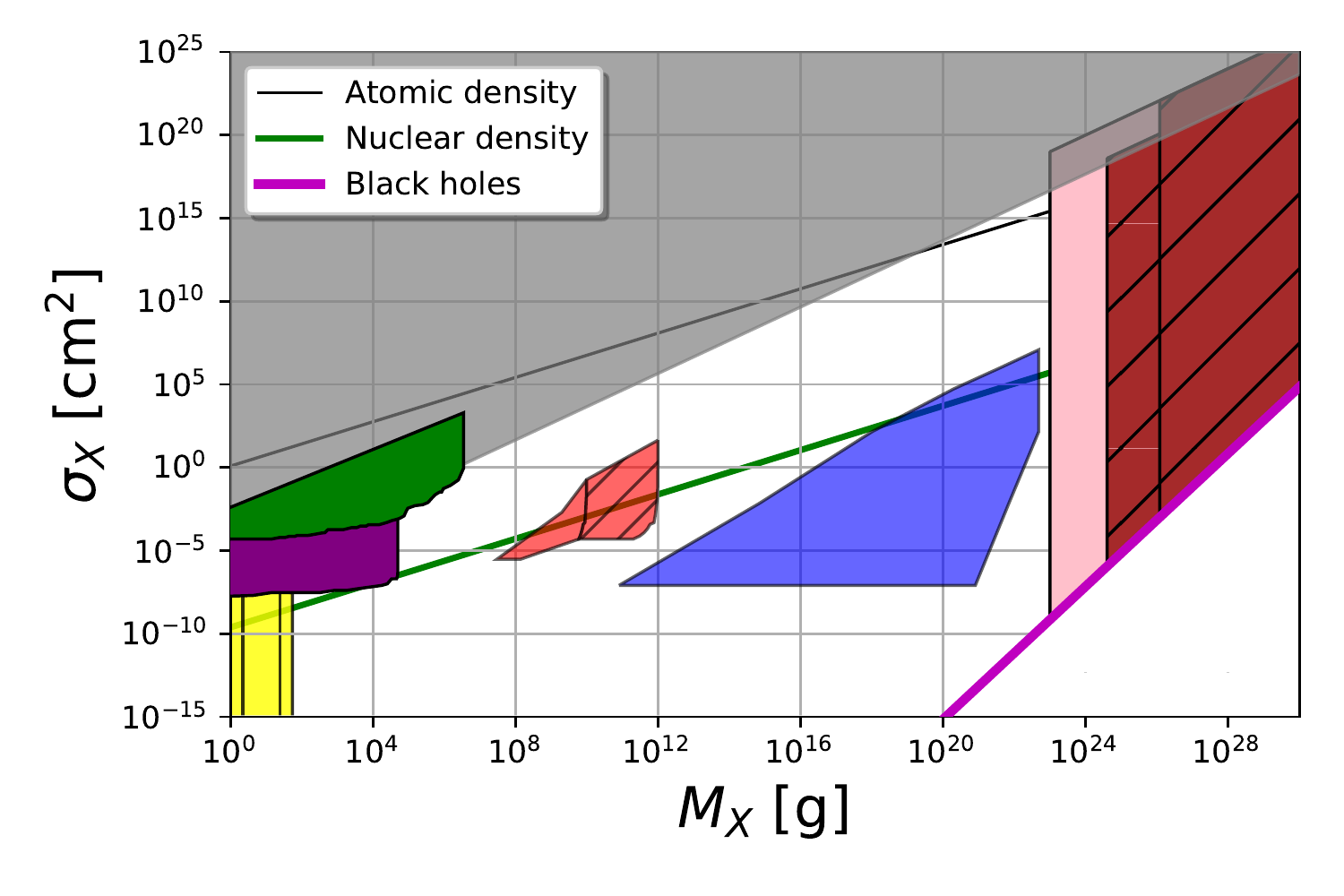}
\caption{Constraints for macros considering elastic scattering to be the dominant interaction. Objects within the region in the bottom-right corner should not exist as they would simply be denser than black holes of the same mass. The various
colored regions represent 
regions where macros cannot
make up the entire amount of dark matter as their
interactions through elastic scattering with their geometric 
cross-section $\sigma_x$ would have
produced observable consequences
that have not been seen.
The purple constraints are derived
from a lack of human impacts
\cite{1907.06674}, the green from a lack of fast-moving bolides
events \cite{1908.00557}, the yellow from mica observation 
\cite{DeRujula:1984axn,Price:1988ge}, the red from superbursts \cite{1912.04053}, the dark blue from white dwarf supernovae \cite{1805.07381}, the grey from structure formation \cite{1309.7588},
the pink from microlensing of M31 \cite{Niikura2019,1910.01285} and
the maroon from microlensing
\cite{Alcock2001,astro-ph/0607207,0912.5297,Griest2013}.
{  See Section I for more details regarding these constraints.}
        }
        \label{fig:exclusion}
 \end{figure*}

There remains a wide range of masses
$M_x$ and cross-sections $\sigma_x$ that
are currently unconstrained by all previous work. 
 
Macros over a wide range of 
densities remain possible candidates to explain
the problem of the nature of dark matter.
The constraints
on macro parameter space
from elastic scattering are
presented in Figure 
\ref{fig:exclusion}.
Some of
these constraints will be relevant to
the results of this paper, as we discuss
below. 

In this work, we introduce an additional
mechanism for energy deposition
through electromagnetic interactions
of charged macros
with charge $Q_x e$ where $e$ is
the unit charge, and $Q_x$
is a number.

The velocity distribution
of macros has in prior works been 
assumed to follow a 
Maxwellian distribution of the form
\begin{equation}\label{maxwellian}
f_{MB}(\bold{v}_x) = 
		\left( \frac{1}{2\pi \sigma^2}\right)^{\frac{3}{2}}
		e^{-\left(\frac{v_x^2}{2\sigma^2}\right)},
\end{equation}
where $\sqrt{2}\sigma \approx 250~ \text{km s}^{-1}$\footnote{
	This is the distribution of macro velocities in a non-orbiting frame moving with the Galaxy.
	When considering the velocity of macros impacting the atmosphere, \eqref{maxwellian} is modified by the motion of the Sun and Earth in that frame, and by the Sun's and Earth's gravitational potential. 
	We have taken into account these effects 
	(as explained, for example, in  \cite{Freese2013}),
	except the  negligible effect 
	of Earth's gravitational potential.
	}. 
We will continue with this assumption
in this work.

Numerous papers have been written considering a variety of charged
dark matter candidates. The vast
majority of the constraints
have been derived for particle dark matter
candidates.
For a review of some
of the constraints
light mass
particles, see e.g. 
\cite{Davidson2000,Dimopoulos1990} and references
therein.
We summarize some of the results 
in the literature
here regarding previous
work on charged dark matter candidates.

A unit electric
charge for a dark matter
candidate, to contribute
the majority of the observed dark 
matter, is excluded if its mass is 
not very large (these excluded
masses are predominantly in the
range of particles masses, but
extend to masses $M_x \leq 
10^{-13}\,$g).
However, as we shall show
dark matter candidates with
a unit charge and much larger
masses, or even charges 
much higher than $\sim e$ 
at much larger masses
are allowed to still
contribute all of the observed
dark matter.

Charged massive particles 
(CHAMPs) with
integer charge values
have been considered (see e.g.
\cite{Dimopoulos1990})
and a variety of terrestrial
and astrophysical constraints
derived. Such bounds
apply only to particles
with $M \lesssim 10^{-13}\,$g.

Millicharged candidates,
i.e. particles with $\epsilon \ll e\,$ where $\epsilon$ is a
fractional charge value, are
constrained by many observations.
Accelerator searches
(e.g. one carried out at SLAC
\cite{Prinz1998} designed specifically to detect millicharged objects) assumed
the millicharged particles to be produced entirely 
via electromagnetic interactions and produced no results 
over the range of sensitivity. 
Constraints have also been
derived from supernova 1987A
\cite{Chang2018},
considering the millicharged particle
to be a dark-sector particle
with a small electric charge.

Others have considered
charged Planck-scale relics
(CPRs) \cite{1906.06348}, which
are expected to be of 
approximately the planck mass
and possess a
charge-to-mass ratio
of $e/M_{pl}$.
Reference \cite{1906.06348}
derived projections
for the maximum 
abundances of CPRs
based on null observations of
a variety of terrestrial
experiments.

{   Concrete charged macro formation models
include macros formed from the mechanisms described for neutral
macros, e.g. thos described in references \cite{LYNN1990186,1005.2124,1912.02813},
that then acquire a charged
by absorbing nuclei during their lifetime. Such a possible
mechanism for absorbing additional nuclei 
is described in reference \cite{1912.02813}.
Indeed, if a macro consists of a bound state of nuclei, it is
plausible that a collision with a nucleus could result in the
absorption of that nucleus, thus increasing the net charge of the
macro. 

However, in this manuscript we will undertake a phenomenological approach
and consider a broad range of parameters $M_x$, $Q_x$ and $\sigma_x$.}
We 
determine the
regions
of parameter space 
where charged macros with 
$M_x \geq 10^{-13}\,$g.
are currently allowed to
be the sole component of dark matter. 
Thus, 
the results presented here in
Figures \ref{fig:money0}-
\ref{fig:money5}, where
we rule out some region of the $Q_x-M_x$ plane, are the regions
of the parameter space
where the
existence of such charged objects
can contribute
only a sub-component
of the dark matter.
We are concerned only with
$M_x \lesssim 10^{23}\,$g,
above which
a variety of microlensing
results \cite{Alcock2001,astro-ph/0607207,0912.5297,Griest2013,Niikura2019} have ruled
out macros as being the dominant
form of dark matter.

Galactic dynamics have been
used to
constrain dark matter
self-interactions. 
Investigations into the 
allowed strength of dark matter 
self-interactions have been
conducted (see e.g 
\cite{Randall2008}).
The observation of an
offset between the gas and dark 
matter in a merging cluster, such as 
1E 0657-56 (a.k.a.
the bullet cluster),
arising because of the ram pressure 
acting on the gas but not the
dark matter has been used to 
constrain dark matter
self-interactions.
Thus, macros
with too high a charge content
would be prohibited.
We discuss the effect
of electrically charged macros on 
galactic dynamics, resulting
in an effect
similar to the dynamical
friction first discussed
by Chandrashekar 
\cite{Chandrasekhar1943}, and then
by Binney and Tremaine \cite{galdyn}.
We derive the analogous 
expressions for the drag force
experienced by a charged macro
travelling through 
a sea of other similarly
charged macros, and relate this
to an effective
cross-section that we use
to constrain the
charged macros
using the self-interaction
constraints.

We show that
charge bounds derived from
the Cosmic
Microwave Background (CMB)
anisotropies
that
were determined for 
objects of much lower mass
\cite{PhysRevD.88.117701}
also
apply to much larger mass
objects. 
Charged particles with
sufficiently strong coupling to 
baryons would participate in the
acoustic oscillations of 
baryon-photon plasma.
This would affect the CMB radiation 
anisotropies in several ways.
Using this idea
together with recent Planck data,
reference \cite{PhysRevD.88.117701}
severely constrained
the charge content of 
dark matter.

Considerations on galaxy cluster 
scale magnetic fields
affecting the charged dark matter
distributions within a cluster has led
to tight bounds being placed on millicharged
dark matter \cite{1602.04009}.
Bounds were obtained
by requiring that the
motion caused by the
randomly oriented magnetic fields 
should not smear out the dark 
matter
distribution governed
by the gravitational interactions and also 
by demanding that the Lorentz force should
not exceed the gravitational force 
in a cluster. We show that
these bounds extend to much more massive
dark matter candidates.

We then set bounds on the
allowed regions
of the charge-mass 
parameter space 
considering some of the
null results
quoted above in Figure 
\ref{fig:exclusion}.
For both the mica and MACRO
results, macros with a sufficiently
large charge content would
have left a detectable track
in either
detector. A phenomenological
law for ions moving 
at speeds of $\beta \sim 10^{-3}$ has 
been determined in \cite{SRIM}
and will be used in deriving
constraints on the allowed charges
of light mass macros
for these two detectors.

Macros incident on a neutron star
would be moving at moderately
relativistic speeds, $\beta \sim
0.7$, and could
potentially trigger
thermonuclear runaway, resulting
in a phenomena known as a 
superburst.
For ions moving quicker than $\beta 
\sim 0.01$, the Bethe 
equation is an accurate description
of the linear energy deposition.
We constrain the charge content
of macros that would have 
otherwise initiated a superburst
in a shorter time than 
observed \cite{1912.04053}.

For all the other constraints
quoted above
in Figure \ref{fig:exclusion}, 
the macro
would be moving at speeds
appropriate to the usage of the
phenomenological fit
in reference \cite{SRIM}. 
However, the threshold linear 
energy deposition for a signal 
is much higher, requiring a large
value of $Q_x$. At 
such large values the 
phenomenological
fit in \cite{SRIM}
is not valid (as will be explained
in more detail in 
section III where the
theory behind the framework
for calculating the linear energy
deposition is reviewed).
For a general review of
the effects of ions 
passing through matter or
for more details on the two frameworks
discussed here, we refer
the reader to reference \cite{SRIM}.

The constraints placed 
in this paper are
from purely phenomenological
observations, independent
of considerations of the binding
energy a macro of a certain density. 
One should consider
only macros that satisfy
\begin{equation}\label{bindingenergy}
\frac{Q_x^2 e^2}{r_x}< E_b\frac{M_x}{m_b},
\end{equation}
where $r_x = \sqrt{\frac{\sigma_x}{\pi}}$,
$E_b$ is the macro binding energy
per baryon and $m_b \sim 938\,$ MeV is the mass
of a baryon. 
However, theoretical
considerations have failed to 
yield a model-independent
formation mechanism for
macros. Hence,
the
binding energy of a macro
cannot be predicted in a
model-independent way and so we ignore 
this consideration.

Since it is unclear
what binding energy macros
would have, we use nuclear binding
energy as a binding energy of potential
interest
when plotting \eqref{bindingenergy}
in our results figures 
(as an equality)
with $E_b = 8\,$MeV, i.e.
the binding energy of iron
peak elements. However this line
is purely for illustrative purposes;
for such large masses, the macro
is much denser than nuclear density
and so it is likely that the
continued existence of such
objects over cosmological
timescales requires binding
energies much higher than 
that corresponding
to nuclear density objects.

We use Gaussian-cgs units
throughout this analysis.
For simplicity, we consider
all macros to be of the same
mass $M_x$ and charge $Q_x$ (as well as geometric cross-section $\sigma_x$).

The rest of this paper is
organized as follows. In Section 
II, we discuss
constraints from
large-scale structure.
In section III, constraints
are obtained from terrestrial observations and the
time duration between back-to-back
superbursts on 4U 1820-30.
In Section IV,
the results are presented, along
with a discussion of their
range of applicability. 
In Section V, we conclude.

\section{Large Scale Structure}

\subsection*{Constraints
from Self-interacting Dark Matter
(SIDM)}

SIDM was initially proposed to solve inconsistencies between 
the cold dark matter (CDM) paradigm predictions and observations
of structures on scales below a few Mpc, including the
missing-satellite problem 
\cite{PhysRevLett.84.3760}. 
The centres of SIDM haloes are expected to have constant-density
isothermal cores that arise as kinetic energy is transmitted from
the hot outer halo inward.
This results 
in a diminished
central density of the
dark matter halo, an idea
first raised in reference
\cite{PhysRevLett.84.3760}.
Such a scenario can
happen if the reduced cross-section of the dark matter candidate,
$\sigma/M$ (valid for any dark matter candidate), 
is large enough for there to be a relatively high probability of
scattering over a time comparable to the age of the halo.

The result of
strong self-interactions is 
an offset between the bullet sub-cluster mass
peak and galactic centroid; the absence of this observation in
the actual cluster provides a limit on $\sigma/M$.
Comparisons were also made between simulations with SIDM and the 
observed
density profiles and substructure 
counts of other observed
clusters, low-surface brightness spiral and dwarf-spheroidal
galaxies in \cite{Rocha2013}.
In both cases, bounds 
on the strength
of the self-interaction
generally
prohibit dark matter
self-interactions with reduced
cross-sections
\begin{equation}\label{SIDMreduced}
\frac{\sigma_{total}}{M_x} \geq 1\,
\frac{cm^2}{gr}\,,
\end{equation}
{  where $\sigma_{total}$ is the total
cross-section for all interaction
mechanisms, e.g. elastic scattering $\sigma_x$
and the Coulomb force. We will consider both
contributions in this manuscript where relevant
although in vast regions of parameter space either
one dominates.}
We use the reduced cross-section
value to place constraints
on the allowed charge values
of macros
that would have altered
galactic dynamics
through strong self-interaction.

Dynamical friction historically refers
to the deceleration of a
massive object moving through
a population of other objects
due to gravitational 
interactions. This effect has been
discussed in
\cite{Chandrasekhar1943, galdyn}.
Here, we consider a similar effect
that arises from a charged macro
moving through a population
of other charged macros.
We show that for
sufficiently high values
of $Q_x$, the self-interaction
between charged macros
would be sufficiently strong
and is thus constrained.
Such strong self-interactions
would result in a situation where
the high velocity macros
located in the outer halo
lose energy to the more slowly
moving macros located near the
center of the galaxy,
resulting in the central
density being diminished
as the macros in the inner
regions
migrate outwards.

The analogous expression
for the deceleration
experienced by a macro
passing through a population 
of macros,
all with the same charge content,
is obtained by redoing the analysis
in Chapter 7 of \cite{galdyn}.
The expression is equivalent
to the original equation except
for the replacement
\begin{equation}
G^2(M+m)m \rightarrow \frac{2Q_x^4e^4}{M_x^2}\,,
\end{equation}
yielding
\begin{equation}\label{drag}
\frac{d\textbf{v}_M}{dt} = -32\pi^2 log (\Lambda^2 + 1)\frac{Q_x^4 e^4}{M_x^2} \frac{\int_0^{v_M} F(\bold{v}_m)v_m^2 dv_m}{v_M^3} \textbf{v}_M\,,
\end{equation}
where 
\begin{equation}
\Lambda = \frac{v_x^2 M_x b_{max}}{2 Q_x^2 e^2}\,,
\end{equation}
$\bold{v}_{M}$ is
the velocity
of the macro under consideration
and
$F(\bold{v}_m)$ is the phase space
number density defined
in \cite{galdyn} and
differs from the velocity distribution \eqref{maxwellian}
by a factor of the number
density $n_x = \rho_{DM}/M_x$
(for macros of a single mass)
\begin{equation}
F(\bold{v}_m) = n_x f(\bold{v}_m)\,.
\end{equation}

For a Maxwellian velocity
distribution
\eqref{maxwellian},
the integration in
\eqref{drag} can be carried out
analytically, yielding
\begin{widetext}
\begin{equation}\label{newdrag}
\frac{d\textbf{v}_M}{dt} =- \frac{8\pi log( \Lambda) Q_x^4 e^4 n_x}{ M_x^2 v_M^3}\left[Erf(X)-\frac{2X}{\sqrt{\pi}}e^{-X^2}\right]\textbf{v}_M\,,
\end{equation}
\end{widetext}
where $X = \frac{v_x}{v_{vir}}$ and
$Erf(X)$ is the error function. Generally,
the quantity in square brackets will be of order $\mathcal{O}(0.1-1)$
provided $v_x$ is not too small,
i.e. for the majority
of macros in the distribution
\eqref{maxwellian}. We make this
simplifying assumption
in the following calculations, i.e.
that the quantity in
the bracket $\sim 0.1$.

A crucial concept 
in this calculation is that
the effect described above 
results in the macros
in the outer halo
experiencing a negative force,
which is essentially a drag
force. Thus, we relate the
drag force derived in 
\eqref{newdrag} to the expression 
for drag first derived by Epstein
\cite{Epstein} for
objects where the physical size,
r,
is significantly smaller
than the average separation 
\begin{equation}\label{mfp}
L = n_x^{-\frac{1}{3}}\,,
\end{equation}
i.e. $L \gg $ r. The Epstein
drag force is
\begin{equation}\label{Epstein}
F_{drag} = \frac{4}{3}\rho_{DM}\sigma_{eff}
\overline{v_x} v_{x,M} = M_x
\frac{dv_M}{dt}\,,
\end{equation}
where $\sigma_{eff}$
is the effective cross-section
due to the Coulomb interactions
between a macro and all other
macros in the population,
$\overline{v_{x}}$ is the mean
speed of the population of macros
and $v_{M}$ is the speed
of the macro under consideration. 
The
effective cross-section
may
be thought of as the equivalent 
geometric
cross-section for
macros to interact, through
scattering,
similarly
to the charged macros interactions
and produce similar
galactic-scale consequences.
We 
take $\overline{v_{x}} = 250\,$km 
s$^{-1}$. For the fast moving
macros
in the galactic population,
assuming $v_{M} = 250\,$km s$^{-1}$
will result in an underestimate
of at most a factor of 
$\mathcal{O}(2-3)$. Thus, for
simplicity, we take
$v_{M} = 250\,$km s$^{-1}$.

The effective reduced 
cross-section can 
be obtained by equating
this drag force to the 
expression for Epstein drag
\eqref{Epstein}
yielding
\begin{equation}
4\pi \frac{e^4}{v_M^4}
\frac{Q_x^4}{M_x^3} = \frac{\sigma_{eff}}{M_x}\,.
\end{equation}
By requiring the effective
reduced cross-section to be greater
than the threshold \eqref{SIDMreduced},
we obtain
\begin{equation}
Q_x \geq 3\times 10^{16}
\left(\frac{M_x}{gr}\right)^{\frac{3}{4}}
\end{equation}
to not be ruled out
by self-interaction
constraints.
This constraint is represented
in purple in the results figures.

{  
We have considered how sufficiently charged macros would
cause deviations from the observed dark matter density profile.
Thus, this constraint depends on how well the dark matter density profile
can be measured. Currently, this quantity is known at best to
to an accuracy of
$ 10 - 50 \%$ \cite{Benito2019}. Thus, we can say with certainty
that for the range of parameter space for which
the constraints apply, charged macros make up at most
a subcomponent of
$\sim 10 - 50\%$ of dark matter.
}
\begin{widetext}

\begin{figure}
    \centering
    \begin{subfigure}[t]{0.45\textwidth}
        \centering
        \includegraphics[width=0.95\linewidth]{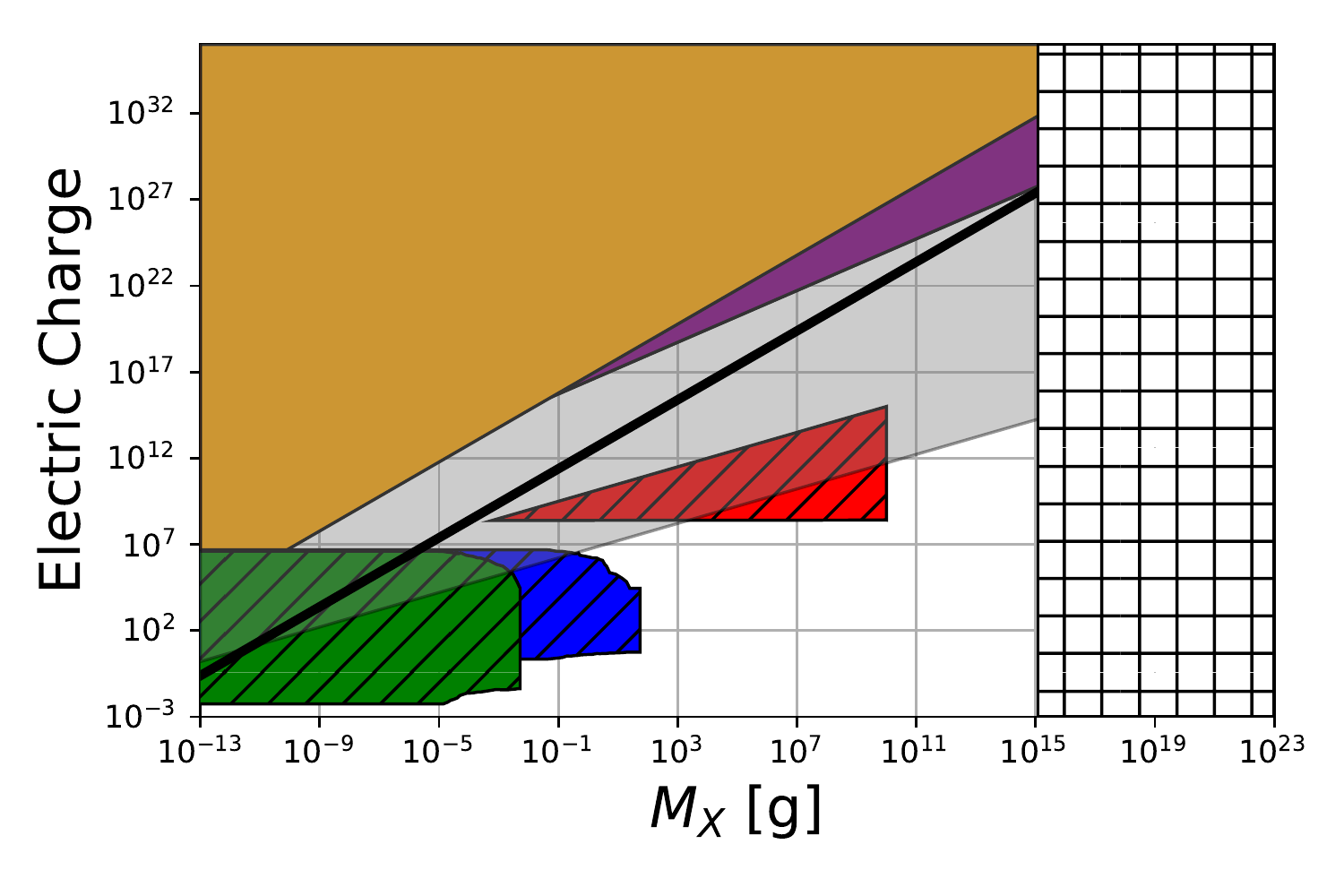} 
        \caption{$\sigma_x = 10^{-25}\,$cm$^2$} \label{fig:money0}
    \end{subfigure}%
    \begin{subfigure}[t]{0.45\textwidth}
        \centering
        \includegraphics[width=0.95\linewidth]{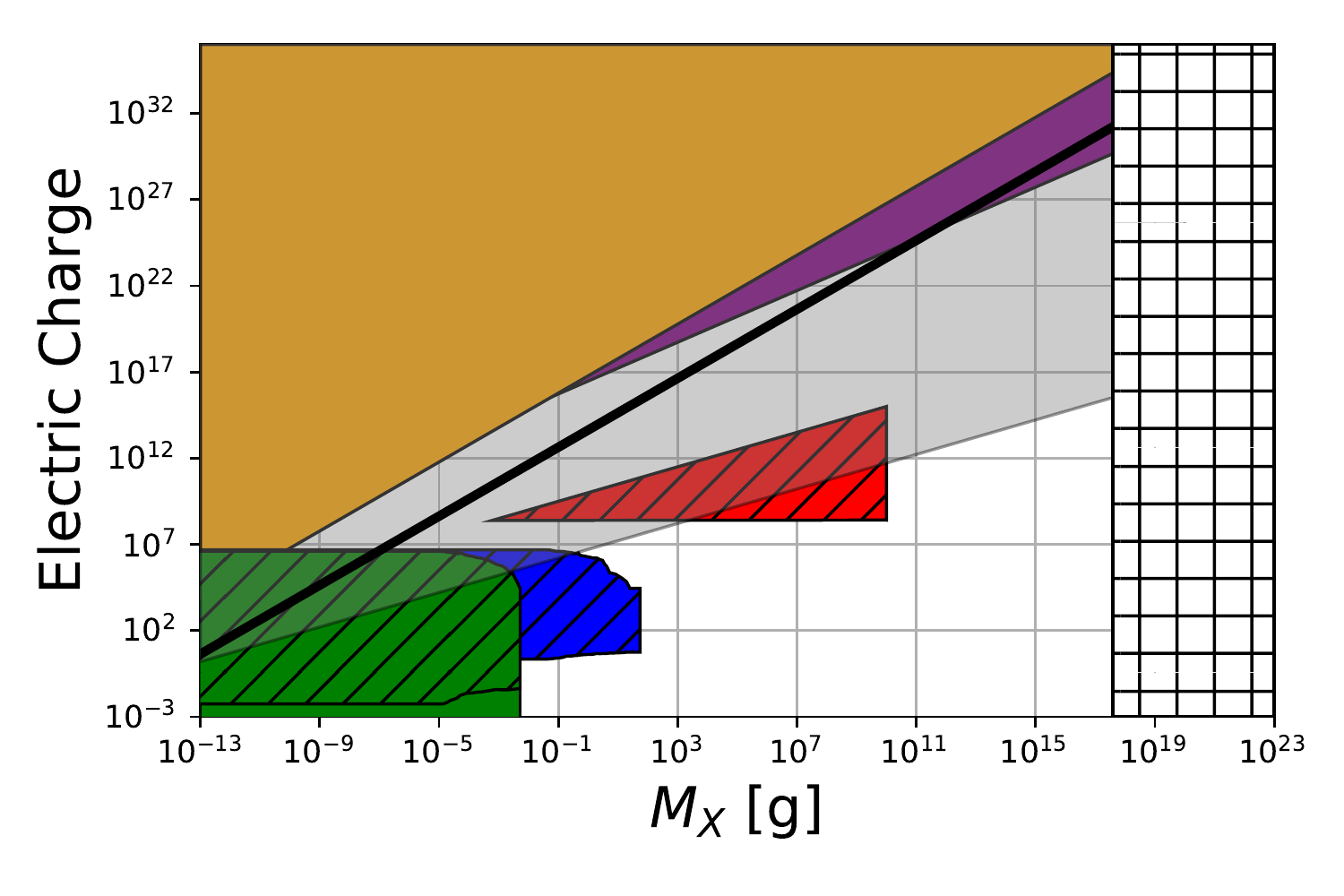} 
        \caption{$\sigma_x = 10^{-20}\,$cm$^2$} \label{fig:money1}
    \end{subfigure}
    \vspace{1cm}
    \begin{subfigure}[t]{0.45\textwidth}
        \centering
        \includegraphics[width=0.95\linewidth]{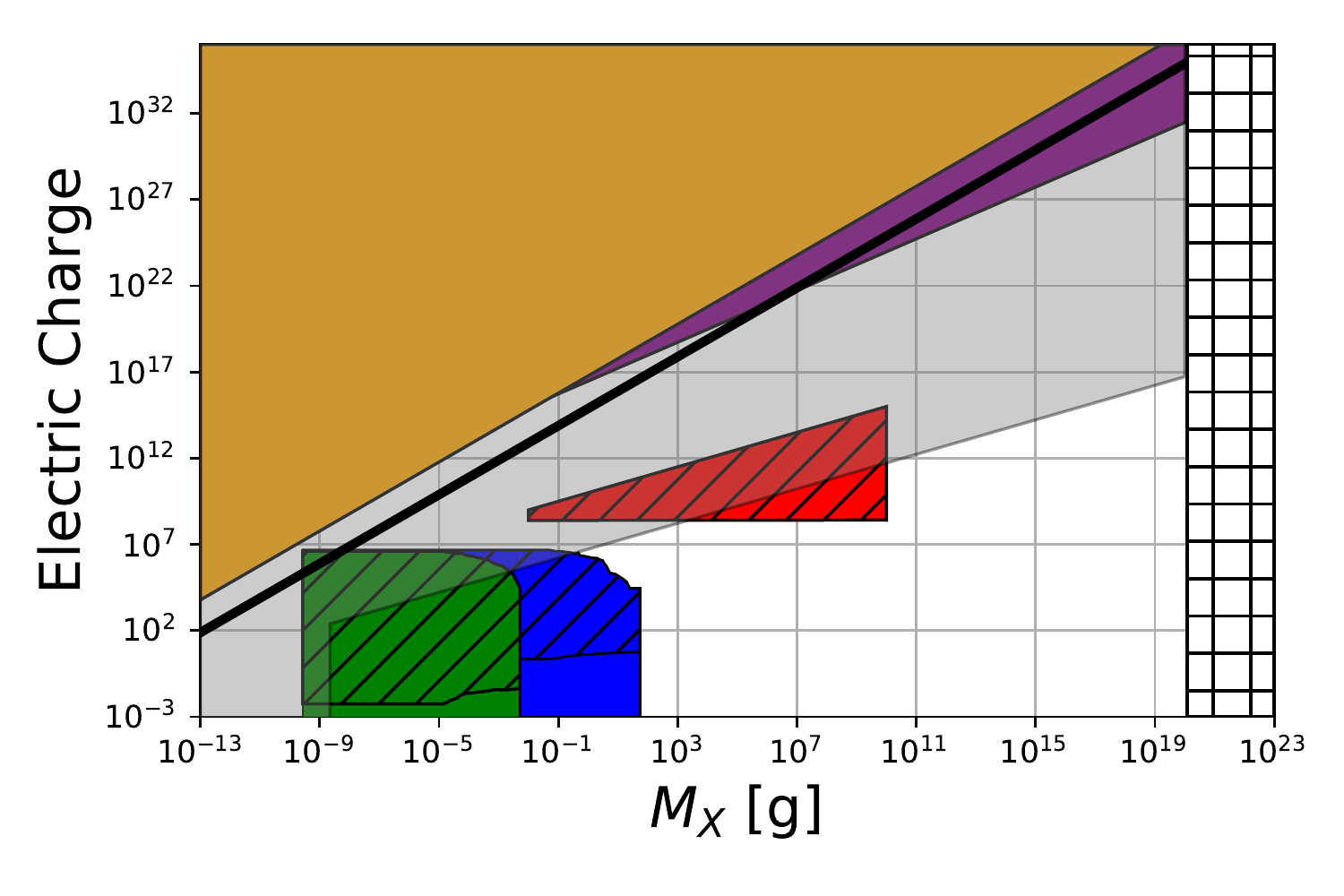} 
        \caption{$\sigma_x = 10^{-15}\,$cm$^2$} \label{fig:money2}
    \end{subfigure}%
    \begin{subfigure}[t]{0.45\textwidth}
        \centering
        \includegraphics[width=0.95\linewidth]{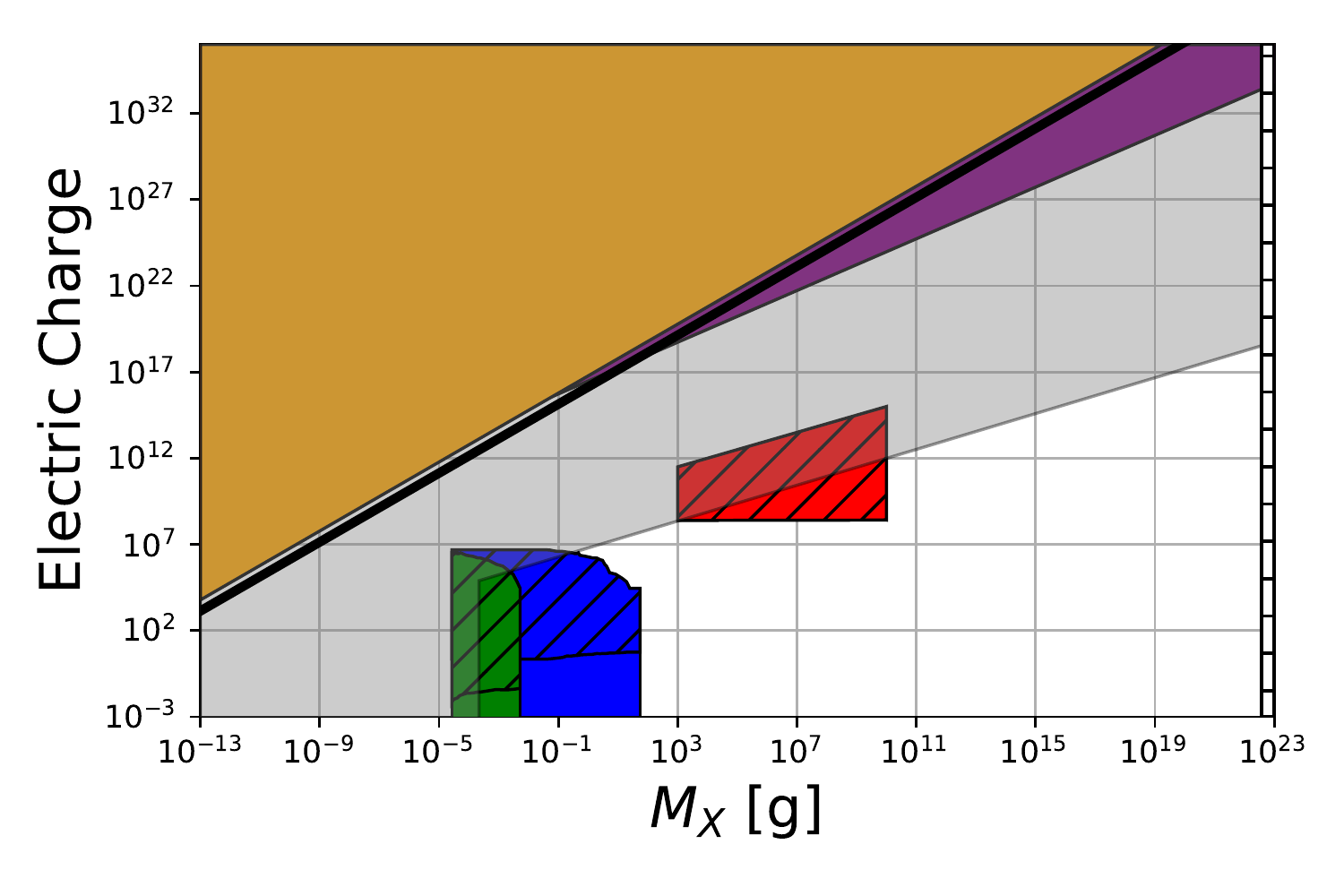} 
        \caption{$\sigma_x = 10^{-10}\,$cm$^2$} \label{fig:money3}
    \end{subfigure}
    \vspace{1cm}
    \begin{subfigure}[t]{0.45\textwidth}
        \centering
        \includegraphics[width=0.95\linewidth]{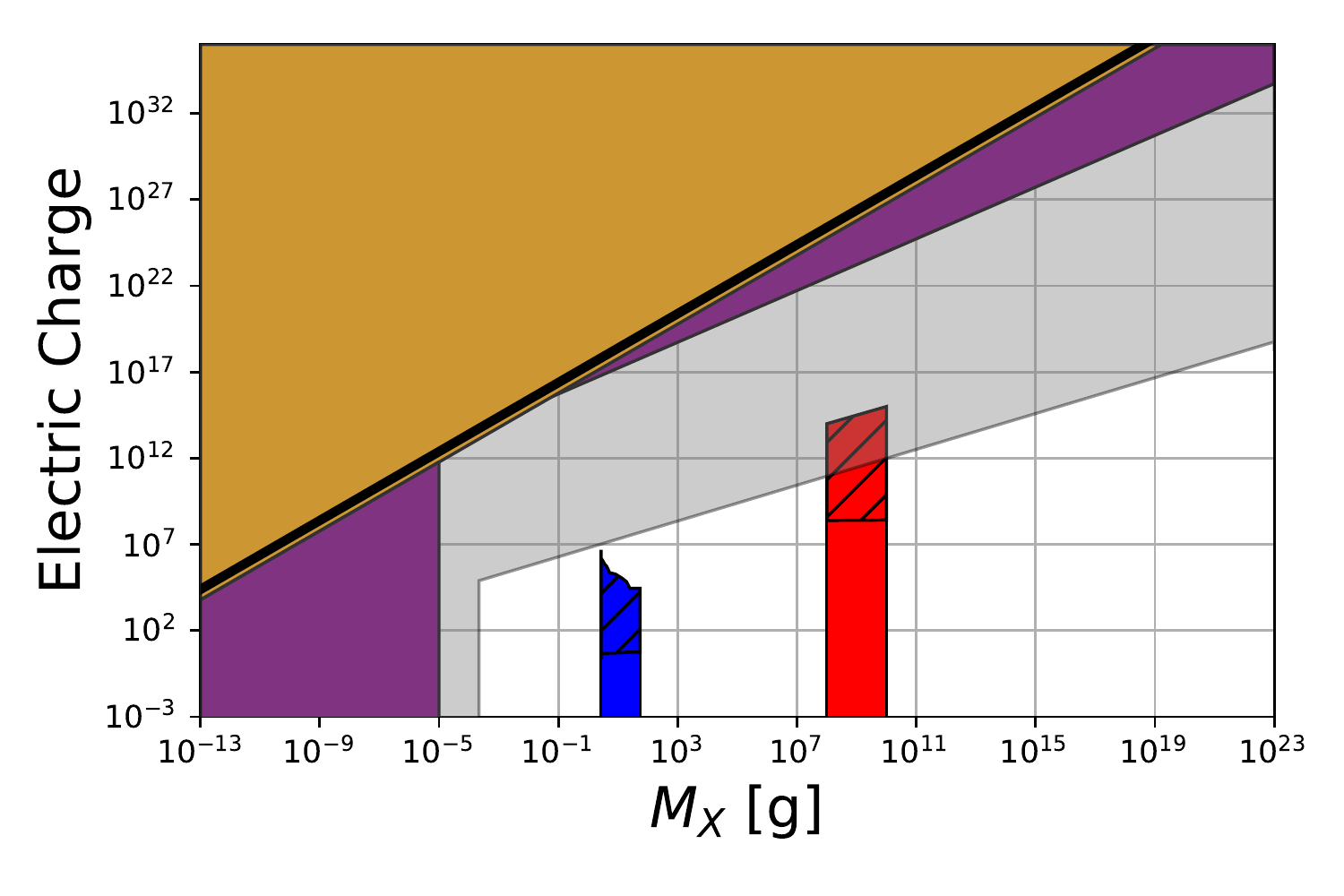} 
        \caption{$\sigma_x = 10^{-5}\,$cm$^2$} \label{fig:money4}
    \end{subfigure}%
    \begin{subfigure}[t]{0.45\textwidth}
        \centering
        \includegraphics[width=0.95\linewidth]{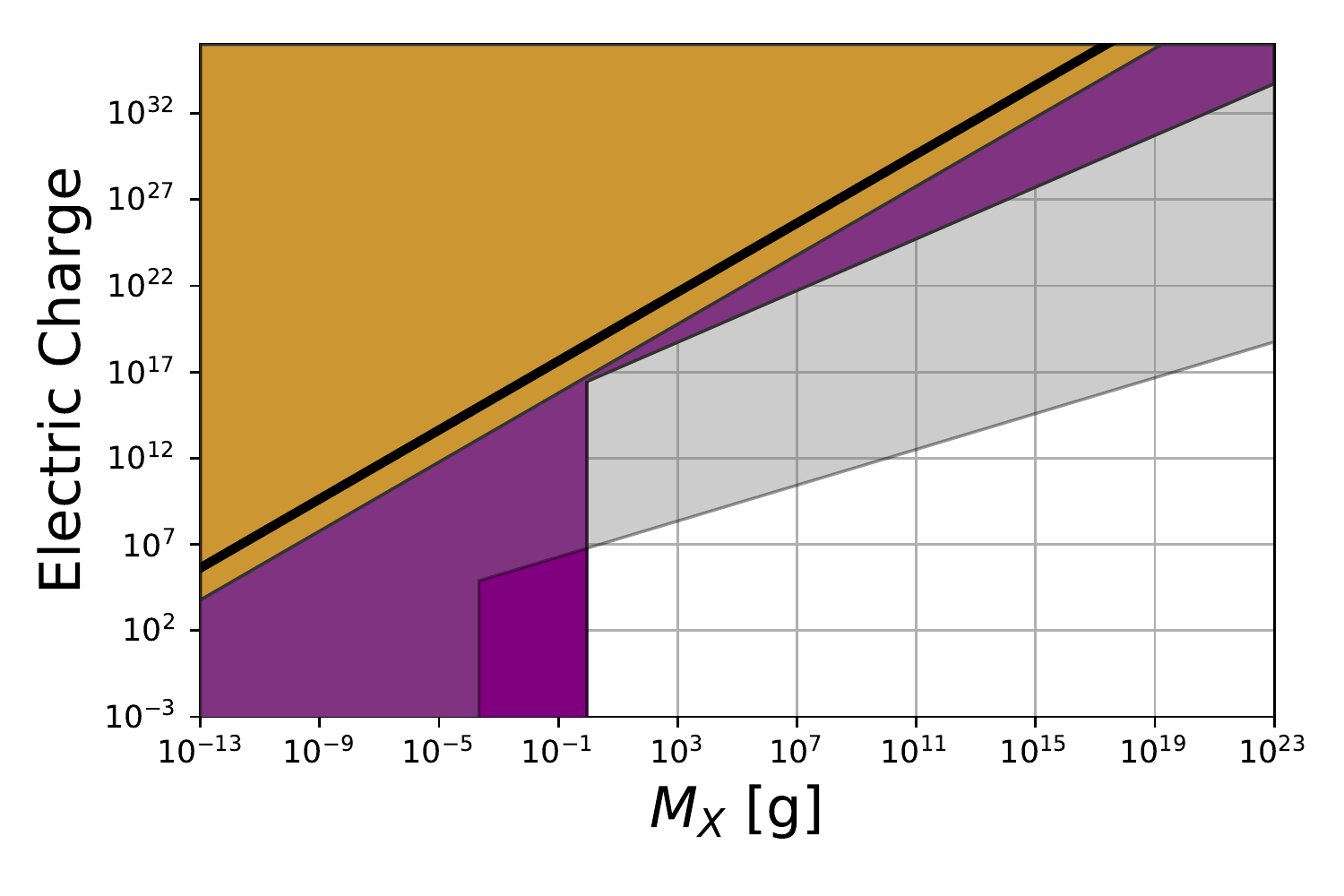} 
        \caption{$\sigma_x = 10^{0}\,$cm$^2$} \label{fig:money5}
    \end{subfigure}%
    \caption{Constraints on the
charge content of macros
for several different values of the geometric (and elastic) cross-section
$\sigma_x$.
Constraints in yellow
are derived from requiring
charged dark matter not alter dynamics
of galaxy clusters \cite{1602.04009}, 
in purple
from requiring charged dark 
matter not interact too strongly
with itself, in grey from
CMB anisotropy considerations
\cite{PhysRevD.88.117701}, in
green from a null result
of the MACRO detector, in blue
from a null result of tracks
in ancient
muscovite mica and in
red from the time
between back-to-back 
superbursts on 4U 1820-30.
The black line represents objects
with binding energy $E_b = 8\,$MeV,
i.e. nuclear binding energy satisfying
\eqref{bindingenergy}.
Objects with masses
greater than a critical threshold \eqref{schwarzchild} should not
exist as they would be denser
than black holes of the same
Schwarzchild radius.
{  The hatching of the mica, MACRO and superbursts
constraints refers to constraints derived in this work
from electromagnetic interactions and not elastic scattering
as in prior work on macros. Those are presented, where relevant,
in the same respective color as the three aforementioned constraints
but with no hatching.}}
\end{figure}
\end{widetext}

 \begin{figure*}
  \includegraphics[width=\textwidth]{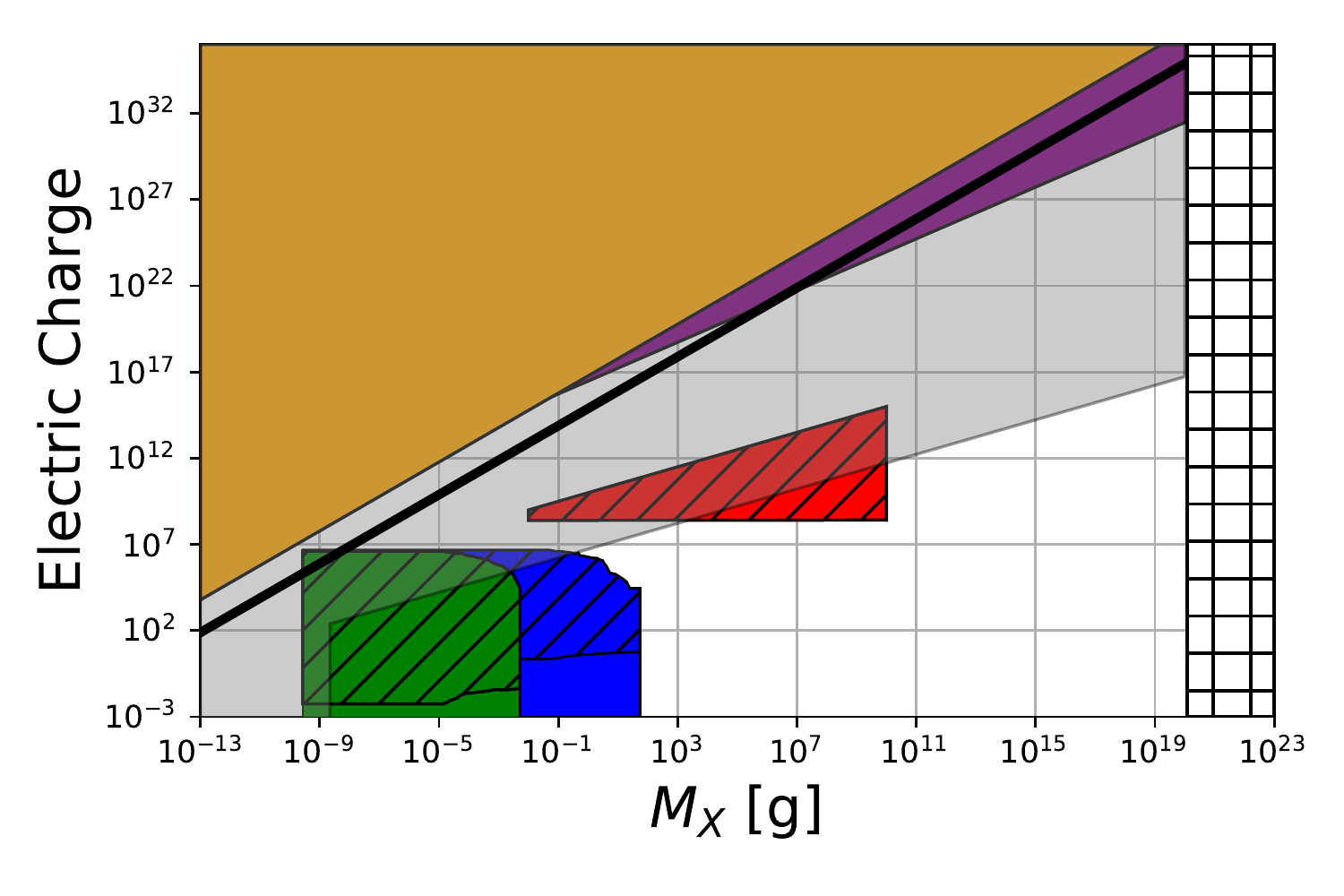}
\caption{Constraints on the
charge content of macros
for geometric (and elastic) cross-section
$\sigma_x = 10^{-15}\,$cm$^2$.
Constraints in yellow
are derived from requiring
charged dark matter not alter dynamics
of galaxy clusters \cite{1602.04009}, in purple
from requiring charged dark 
matter not interact too strongly
with itself, in grey from
CMB anisotropy considerations
\cite{PhysRevD.88.117701}, in
green from a null result
of the MACRO detector, in blue
from a null result of tracks
in ancient
muscovite mica and in
red from the time
between back-to-back 
superbursts on 4U 1820-30.
The black line represents objects
with binding energy $E_b = 8\,$MeV,
i.e. nuclear binding energy satisfying
\eqref{bindingenergy}.
Objects with masses
greater than $M_x \sim 1.2 \times 10^{20}\,$g should not
exist as they would be denser
than black holes of the same
Schwarzchild radius.
        }
        \label{fig:moneytest}
 \end{figure*} 
 
\subsection*{CMB constraints}
Reference 
\cite{PhysRevD.88.117701}
derived bounds on
the charge of
millicharged
particles based
on CMB anisotropy
measurements and using
data from Planck.
However, charged dark matter objects,
regardless of their mass,
scatter off electrons and photons
at the epoch of 
recombination.
It was shown 
\cite{PhysRevD.88.117701} 
that if the
velocity transfer rate of this process exceeds the expansion
rate of the Universe, the millicharged particles behave
similarly to baryons until recombination.

This was used to constrain the
charge content of
millicharged particles.
This bound may be restated as
\begin{equation}
Q_x \geq 5 \times 10^6
\left(\frac{M_x}{gr}\right)^{\frac{1}{2}}\,.
\end{equation}

To justify the application of this 
bound to
charged macros, which can be much
larger than even Planck mass
objects, the diffusion time,
$t_{diff}$, for a photon to cross
this average separation
must be short compared to the 
relevant Hubble time.
We assume that the Hubble time
is that which corresponds
to a radiation dominated Universe
for simplicity
\begin{equation}
H^2 = H_0^2(1+z)^4\,,
\end{equation}
where $H_0 \sim 70\,$km s$^{-1}\,$
Mpc$^{-1}$ \cite{1908.04619} 
is the value of the
Hubble constant today.

The 
average macro separation is
determined from the number
density of a distribution of macros
of a single mass \eqref{mfp}.
The diffusion time for
a photon, interacting with
a mean free path
\begin{equation}
\lambda_{MFP} = (n_e \sigma_T)^{-1}\,,
\end{equation}
where $n_e$ is the electron
number density and $\sigma_T$
is the Thomson scattering
cross-section, is
\begin{equation}\label{tdiff}
\tau_{diff} = \frac{L_{DM}^2}{\lambda_{MFP} c}\,,
\end{equation}
where the quantity
in the denominator is
the thermal diffusivity.

Requiring \eqref{tdiff}
to be small compared to the Hubble time, H$^{-1}$
yields
\begin{equation}
n_e \sigma_T n_x^{-\frac{2}{3}}c^{-1}
H \ll 1
\end{equation}
In terms of the Macro mass this inequality may
be written as approximately
\begin{equation}
    \left(\frac{M_x}{gr}\right)^{\frac{2}{3}}
    (1+z)^3 \ll 10^{39}.
\end{equation}
For the large Macro candidates of 
$M_x \sim 10^{25}\,$g, i.e.
much greater than masses
we are interested in probing,
$t_{diff,DM} \leq
H^{-1}$
remains true for z $\geq 10^7$.
Thus, the large dark
matter masses considered
here do not
ruin the dark matter fluid
approximation and  
the constraints originally derived
for particle mass dark matter
applies equally to macros
at redshift $z \sim 1100$
when recombination took place.

However, the constraints
from the CMB differ from
all other constraints in this
manuscript in one way:
the CMB constraints
are from the early Universe
while all other constraints
are from the late Universe.
We would require
some formation mechanism
capable of producing charged 
macros in the required
abundance
by the era of recombination
to consider the CMB
constraints
on the same footing
as the other constraints.
Thus, we present the CMB constraints
in the results figures
in light shading to
show that the CMB constraints
are subject to additional
scrutiny.

{  
The original constraints limited the abundance of 
millicharged dark matter to $0.2 \%$ over the range of
applicability of the constraint. This result thus
also applies to charged macros.}

\subsection*{Large Scale Magnetic
Fields}

A stringent bound was placed 
on the charge content
of dark matter
in \cite{1602.04009} using
magnetic fields in galaxy clusters.
Since
magnetic fields of $ B \sim 1\, \mu 
G$ typically exist in clusters, upper 
bounds on the charge of dark matter 
were
derived by looking into the effects 
of the magnetic
on the charged dark matter.

In particular, the constraints were
derived by
requiring that the motion 
induced by the
magnetic fields 
should not change the charged
dark matter distribution 
governed
by the gravitational interactions.
Similar constraints were derived
by requiring the Lorentz force 
not exceed the gravitational force in 
a cluster, since
dark matter
interacts predominantly
through gravity
on such scales.
Charged dark matter
with 
\begin{equation}
Q_x \geq 10^{16}\left(\frac{M_x}{g}\right)
\end{equation}
were ruled out based on this analysis.

To 
justify the application of this 
bound to
macros over the mass range of 
interest, which is much larger
than the masses considered in 
reference \cite{1602.04009}, 
we must justify the dark
matter fluid approximation.
We
must demonstrate that the 
physical volumes
considered are larger than 
$n_x^{-1}$ since we are using
the distribution of the dark matter
to place constraints.
Given an average cosmological dark matter 
density of 
$2 \times 10^{-30}\,$
g cm$^{−3}$
and a
maximum macro mass of interest 
$10^{21}\,$g, we must consider
comoving volumes greater than
of order 1 pc
which is true of
the probe considered here.

This bound is stronger
than the bound from
self-interactions for
$M_x \lesssim 1\,$gr.

{ 
Similar to the constraint derived from self-interactions,
this constraint also depends on how well the dark matter density
profile is known and so the same maximum abundance of charged macros
as dark matter can be inferred for this constraint.
}

\section{Energy 
Deposition along Tracks
of charged objects}

In this section, we 
are concerned with
localized terrestrial
and astrophysical detectors
where the 
linear energy deposition
of a passing macro would
exceed some critical
threshold and leave an
observable signal.
The linear energy 
deposition is now the sum
of the two separate
contributions
\begin{equation}\label{totaldedx}
\frac{dE}{dx} = 
\frac{dE}{dx}\bigg\rvert_{elastic}
+
\frac{dE}{dx}\bigg\rvert_{Coulomb}\,.
\end{equation}
{ 
We are most interested in cases where the second term
alone exceeds the threshold energy deposition
for a track to be produced in our detector. However,
we will also consider cases where the first term dominates, 
i.e. cases where a large but neutral macro would
have triggered the detector through elastic scattering.
}

As discussed in \cite{PDG,SRIM},
the speed of a passing ion
determines the amount of energy transferred on passing through
a material. This
is primarily because 
different energy
transfer mechanisms
dominate at different speeds
(see additionally e.g. 
\cite{hellborg_2011,9789048145102}
for a discussion
of the differences between electronic
and nuclear stopping). 
We first discuss
macros moving slowly $\beta \sim 10^{-3}$ before proceeding
to
the moderately 
relativistic version
$\beta \sim 0.7$.
The form of the second
linear energy deposition term
in \eqref{totaldedx}
will be different for
each of these cases.

\subsection*{Non-relativistic
macros}
 
Reference \cite{SRIM}
has produced 
a reasonably accurate model
of the energy transfer of 
low energy
ions by taking into account
the complex electronic
screening potential.
The goal of reference \cite{SRIM}
was to produce a
single analytic function
for the interatomic potential,
thus
allowing a single formula
for the nuclear stopping to be 
determined, as opposed
to using a separate function
for each ion-atom pair.

The potential of the two
particles can also be reduced
to that of a single potential 
called the Interatomic Potential
\cite{SRIM}.
Each of these potentials may
be considered as a Coulombic
term multiplied by a
screening function,
due to electronic screening
that reduces the effects
of the nuclear Coulombic term
at all radii.
The screening function
is related to the 
interatomic potential through
\begin{equation}
\Phi = \frac{V(r)}{Z_1 Z_2 e^2/r}\,,
\end{equation}
where $Z_1$ and $Z_2$ are the 
bare charge of the two
interacting nuclei (in our case
we take $Z_1 = Q_x$) and $r$ is
the distance between the nuclei.
The interatomic function
is generally found by using
simple atomic potentials and
adjusting the definition
of the screening length
to approximate the
two-atom potential.

To obtain one analytic
function to describe
nuclear stopping in all
ion-atom pairs,
a large sample of ion-atoms
pairs were chosen and the
detailed potential calculated 
using
computer simulations \cite{SRIM}.
Various
screened potentials of the form
\begin{equation}
\Phi(x) = f(x)\,,
\end{equation}
where
$x = \frac{r}{a_{screen}}$
and $a_{screen}$ is a 
screening length, which is
a parameter that
characterizes the radial 
spread of the electronic
charge about the radius
\cite{SRIM},
were then trialed to determine
the screening potential
that most
closely matched all
the pairs. The 
screening function
that was determined to
best fit the pairs 
\begin{widetext}
\begin{equation}
\Phi(x) = 0.1818 e^{-3.2x} +0.5099e^{-0.9423x} +0.2802e^{-0.4028x}
+0.02817e^{-0.2016x}\,
\end{equation}
\end{widetext}
had a screening length that was determined to be
\begin{equation}\label{screeninglength}
a_u = 0.8853 a_0 \frac{1}{Q_x^{0.23} + Z_2^{0.23}}\,,
\end{equation}
where $a_0$ is the Bohr radius. 
This quantity
was determined to
be the one with
the appropriate $Z$ dependence
to best reproduce the results
of the numerically calculated
potentials
from the 522 atomic pairs.
The screened 
potential was
determined to deviate from
experimental measurements
by at most $18\%$ \cite{SRIM},
which is sufficient
for the purpose of this
manuscript.

With the
universal screening potential,
the energy transferred due to
the two scattering of 
the two particles
can be calculated
as
\begin{equation}\label{stopping}
S_n(E) = \int_0^\infty T d\sigma\,,
\end{equation}
where T is the energy transfer
by the passing macro
and $d\sigma = 2\pi b\, db$.
The stopping power in 
\eqref{stopping}
is related to the linear 
energy deposition through
\begin{equation}
\frac{dE}{dx}\bigg\rvert_{Coulomb} = N S_n(E)\,,
\end{equation}
where $N$ is the number density
of atoms.

Results are generally presented
in terms of a reduced energy,
$\epsilon$, and a corresponding
stopping power $S(\epsilon)$.
These are related to the
physical versions of
these quantities through
\begin{equation}
\epsilon = \frac{a_u E_0 M_2}{Q_x Z_2 e^2 (M_x + M_2)}\,,
\end{equation}
and
\begin{equation}
S(\epsilon) = \frac{\epsilon}{\pi a_u^2 \gamma E_0}S(E_0)\,,
\end{equation}
where $a_u$ is the universal
screening length
in \eqref{screeninglength}.
The purpose for converting
to a reduced coordinate
system was to better show the
results of using classical
charge distributions
and solid state distributions,
which was also done for the
first time in the 
calculations of \cite{SRIM}.
In such a coordinate system,
a single curve describes
all combination of atom-atom
collisions.

For ease of calculation, 
an analytic fit to the solution
of the reduced stopping power
was given
\begin{equation}\label{fit}
S_n(\epsilon) = log \frac{(1+a\epsilon)}{\epsilon + 
b \epsilon^c + d \epsilon^e}\,,
\end{equation}
where the best-fit coefficients were determined to be
$a = 1.1383$, $b = 0.01321$,
$c=0.21226$, $d = 0.19593$ and
$e = 0.5$. 

This function 
is related back
to the physical stopping power
through
\begin{equation}
S_n(E_0) = \frac{8.462 \times 10^{-15}Z_1Z_2M_2S_n(\epsilon)}
{M_1 + M_2 (Z_1^{0.23} + Z_2^{0.23}}  \frac{eV}{\frac{atom}{cm^2}}.
\end{equation}
It is this function
that we have used in determining
the minimum value of $Q_x$ for
a macro to have a left a detectable
track in the MACRO experiment and
slab of mica, together with an
approximate value of $N \sim 10^{23}\,$atoms cm$^{-3}$ in
both cases. 

This fitting procedure is valid
only for $\epsilon \gtrapprox 10^{-5}$, below which larger charges
produce smaller energy depositions.
This can be seen by taking the
low $\epsilon$ limit
of \eqref{fit}, the middle term in
the denominator dominates
for $\epsilon \gtrapprox 10^{-5}$.
Thus, \eqref{fit} becomes
\begin{equation}\label{weird}
S_n(\epsilon) \propto \epsilon^{0.78774}\,,
\end{equation}
which results in 
\begin{equation}\label{weird2}
S_n(E) \propto Q_x^{-0.01774}\,.
\end{equation}
Thus, larger values of $Q_x$
reduce the energy transfered to
the surrounding medium
and we truncate our analysis
once the value of $\epsilon \sim 
10^{-5}$ because one would
expect that larger values
of $Q_x$ would deposit more energy
in the surrounding medium. The
phenomenological law breaks down in this
regime of $\epsilon \leq 10^{-5}$.

The procedure described in
\cite{SRIM} was performed
using 522 pairs of atom. 
Experimental verification
has been conducted
using various
ion-atom pairs. However,
the results derived here
will be at charge values far
above those tested and
verified. It is thus
quite
possible that we are extending
the results
of \cite{SRIM} into
a region of parameter space 
where it is not an accurate
description of interactions
between ions and atoms. Nonetheless,
it is reasonable to suggest
that at such large charge values,
energy deposition would indeed
be high. Thus, although the 
tools used may not be accurate,
these tools are 
currently the best tools
available and we utilize
them to the full extent
permitted.

The preceding discussion is
relevant to both the MACRO
detector, and ancient muscovite
mica, which we discuss next.

\subsubsection*{MACRO and mica}
Macros of a sufficiently
low mass would have left
an observable signature
on Earth. If they have a low
enough $\sigma_x/M_x$ and charge 
value $Q_x$ so that they 
would have penetrated deep (a few km)
into the Earth’s crust, a 
record would have been
left in the MACRO
experiment and ancient muscovite 
mica. 
We will use the lack of a track
in both these detectors
to constrain the charge content
of macros of low masses.

MACRO was a large multipurpose underground detector located in the Hall B of the Laboratori Nazionali
del Gran Sasso (Italy); it was optimized for the search
of GUT magnetic monopoles with velocity 
$\beta \geq 4 \times 10^{−5}$ 
\cite{hepex0207020}.
A track
that would have been detectable
by etching measurements would
have been left by macros that
deposited 
a minimum nuclear component of stopping power 
\begin{equation}\label{macro}
\frac{dE}{dx} \sim 5 \frac{\textrm{MeV}} {\textrm{cm}}\,.
\end{equation}
However, the MACRO experiment
obtained a null result
and due to the extreme sensitivity
of the detector, constraints
were placed on extremely small
macros from elastic scattering
in \cite{1805.07381}.

Similarly, ancient muscovite mica
was used to constrain macro
parameter space 
\cite{jacobs2015macro}
based on the null
result of tracks when an etching
technique was applied to look
originally for lattice
defects produced by passing
magnetic monopoles
predicted by Grand Unified Theories
\cite{DeRujula:1984axn,Price:1988ge}.
A track would have been
left by macros with
a linear energy deposition
\cite{jacobs2015macro}
\begin{equation}\label{mica}
\frac{dE}{dx} \sim 10 \frac{\textrm{GeV}}{\textrm{cm}}\,.
\end{equation}
These thresholds of
linear energy deposition will
be used in this manuscript
to constrain the charge values
of macros that would
have left a track
independent of the 
geometric cross-section
$\sigma_x$.

\subsubsection*{Lack
of constraints for other
non-relativistic macro scenarios}
In Figure 1, there exist
constraints on the abundance
of macros from numerous other
observations including
the continued existence of white
dwarfs, the lack of fast-moving
bolides in meteorite surveys
and a lack of unexplained human
deaths. 

However, in all cases other than
that of MACRO and mica, 
we are unable to derive
a minimum possible value of $Q_x$.
Hence, we are also unable to
determine the maximum $Q_x$ for a 
charged macro to not lose
most of its momentum and stop
before reaching the appropriate
depth. 

We are unable to determine
a lower bound
on $Q_x$ because
the amount of charge for 
a macro to be capable
of producing 
any of the aforementioned events
is too large
as a much higher
threshold linear energy deposition
is required
than either \eqref{macro}
or \eqref{mica}.
For such large 
$Q_x$ values, the phenomenological
law breaks down as it enters
a region where it's validity
is questioned. Such large values
of $Q_x$ result in $\epsilon \lesssim 10^{-5}$, where the theory
results in predictions that are 
counter-intuitive (see discussion
around equations \eqref{weird}and \eqref{weird2}). Thus
although it seems likely that
some range of
charges might be constrained
by these observations,
there is currently
no theory capable 
of rigorously predicting
this range. Thus, we do not
use any of the other constraints
from $\sigma_x$
to place any constraints
on the macro charge $Q_x$.

\subsection*{
Moderately 
Relativistic macros}
For moderately relativistic charged heavy particles,
the energy loss is well described
by the Bethe equation \cite{PDG}. 
Classically, the derivation
by Bohr assumed the electrons
were stationary. The quantum
mechanical version was later
derived by Bethe
and does not deviate significantly
from the classical version
where we are concerned. 
For the purpose of this section,
we are concerned with the injection
of a large amount of energy 
into ions
near the surface of a neutron star
with the purpose
of triggering thermonuclear runaway 
resulting in a superburst. Thus, 
electronic corrections
as in the shell-corrections,
are not required.
The density
effect correction is also unimportant
at such low (but still
moderately relativistic) speeds 
\cite{PDG}.

Although one expects
for moderately 
relativistic macros that the electronic
energy transfer is stronger
than the nuclear component,
as mentioned above, we 
are interested in the carbon
ions near the surface
of a neutron star that
can undergo thermonuclear
runaway. 

Considering a macro
through the outer layer
of a neutron star, the net
momentum transfer experienced by the
ions perpendicular to the
direction of motion
of the passing macro
due to the Coulomb
force,
\begin{equation}\label{E}
E_{per} = \frac{Q_x e b}{(b^2 + (vt)^2)^{\frac{3}{2}}}\,,
\end{equation}
is given as
\begin{equation}\label{momentumtrans}
p = \int_{-\infty}^{\infty}
dt F_{per} = 2 \frac{Z_1 Q_x e^2}{b v}\,.
\end{equation}
In the non-relativistic limit
where the energy transferred
can be well approximated as
just the classical component,
the energy transferred is
\begin{equation}
\Delta E = \frac{p^2}{2m}
= 2 \frac{Z_1^2 Z_2^2 e^4}{b^2 v_x^2 m_c}\,.
\end{equation}

To obtain the stopping power, 
$S_n(E)$, this energy transferred
must be integrated over
all impact parameters
\begin{equation}
S_n(E) = 2\pi \int \Delta E(b)
b db\,
\end{equation}
resulting in a linear energy 
deposition 
\begin{equation}\label{Bethe}
\frac{dE}{dx} = \frac{4 \pi N Z_1^2 
Q_x^2 e^4} {{m_C v_x^2}}
log \left(\frac{b_{max}}{b_{min}}\right)\,,
\end{equation}
where $N$ is the number density
of atoms in the medium of the
detector, $Q_x$ is the charge
of the macro, $m_C = 10^{-23}\,$g
is the mass of a carbon nucleus and
$b_{max}$ and $b_{min}$ are the
upper and lower limits of
integration.

This is the classical form
of the Bethe equation
first derived by Bohr \cite{SRIM}
and is sufficient for
our purposes due to the
simplifications mentioned previously.

To determine
the limits of integration, we first
summarize the theory behind
thermonuclear runaway.
As discussed in \cite{1505.04444,1805.07381},
for thermonuclear runaway
to be ignited, there is 
a minimum sized region ($\lambda_{trig}$)
that must be raised above a threshold
temperature $T_{crit} \sim 5\times 10^9\,$K for thermonuclear
runaway to be initiated. 
($\lambda_{trig}$) is strongly
dependent on density. Thus,
the upper limit is
the trigger size.
The lower limit
in the logarithm is taken
to be the physical
size of the macro
nuclei, i.e. 
the nuclei that are impacted head on
by the macro (since by definition
we are considering macros
who elastic scattering cross-section
is below the minimum size necessary
to trigger thermonuclear runaway)
are not important for the purposes of this manuscript. All nuclei around these central ones are of interest.

However,
since the limits are only present
in \eqref{Bethe} inside
the logarithm, the
results derived here will
be relatively insensitive
to those limits. For
the range of trigger sizes
determined in 
\cite{Timmes1992,1505.04444},
the logarithm gives a factor
$\sim 10$, and we will
use this approximation
to simplify the analysis.
\subsubsection*{4U 1820-30}
We use the time between back-to-back
superbursts on a neutron star
4U 1820-30
to determine constraints
on the charge content of  dark 
matter of higher masses
than previously constrained
with terrestrial detectors.

A macro passing through 
4U 1820-30 would have set off
a superburst provided a
linear energy deposition
of
\begin{equation}
\frac{dE}{dx} \gtrapprox 
6\times 10^{22} \frac{MeV}{cm}\,,
\end{equation}
had been deposited. 
This would have resulted
in $\sim 10^{8}\,$J of energy being
deposited over a range of
$10^{-4}\,$cm, which was
the trigger size for
a density of $\rho \sim 10^8\,$
g cm$^{-3}$ (insert citation here).
For less dense regions in a neutron
star crust, the energy requirement is
higher but the general process 
for determining the threshold
linear energy deposition
is the same.

Accreting neutron stars undergo
superbursts naturally once
enough material has been 
accreted \cite{1912.04053}.
However, a macro incident on 
such a neutron star could trigger
a superburst, which would 
not be as powerful as
one caused by no external
trigger. This idea was used
to constrain intermediate mass
macros based on the decade
long duration between back-to-back
superbursts on 4U 1820-30.
However, there is one caveat to this
\cite{1912.04053}
constraint as it is still
unclear how superbursts are
initiated. This is similar to the case
of white dwarfs undergoing a type 1A
supernova. It is unclear if the
initiation of a deflagration wavefront
is sufficient to trigger
thermonuclear runaway in the entire
carbon ocean. This will require
further numerical work to determine
if the region constrained is truly
ruled out.

{ 
The constraints derived in this section all depend primarily on
the flux of charged macros, i.e. they are proportional to $M_x^{-1}$.
Thus the abundance of charged macros can be constrained as $M_x^{-1}$.
This implies that the lower the charged macro mass
constrained by these methods, the stronger is the limit
on the maximum abundance of these objects as dark matter. Thus,
smaller mass charged macros can contribute only as tiny
subcomponents of dark matter.
}

\section{Results}
{ 
We first summarize the main parameter(s) that determine
the constraints derived in this manuscript before discussing
other aspects of these constraints.

The strongest constraint for millicharged dark matter (and hence
charged macros) is the CMB constraints,
which is an early Universe constraint and requires the dark matter 
to have been formed
by this point (if not much earlier). The other two
large scale structure constraints are both late Universe constraints. 
In general, millicharged dark matter constraints 
depends on both the charge $Q_x$ and $M_x$. This is a reflection
of the fact that these constraints depend on the charged macros
in the distribution creating large scale effects. This means that what
matters is the amount of ``charge per unit mass". This is not an exact statement
because the constraints depend on the charge per unit mass raised to various powers.
However, this statement is schematically true.

The constraints from ancient mica, MACRO monopole searches and superbursts 
are dependent primarily
on the flux of incident charged macros, i.e. the mass. Additionally,
using the phenomenological law derived in reference \cite{SRIM} that
was used to constrain charged macros utilizing null results
from ancient mica and the MACRO monopole detector, 
the speed of the charged macros in the distribution
determines the upper and lower bounds of the charge values that are constrained.}

\subsection*{Applicability}
Before we discuss the results,
we first discuss the range of 
applicability of the results.

We first consider
macros incident on either
MACRO or mica as well as
the outer layers of a neutron star. 
Considering
elastic scattering alone, macros of a sufficiently
large cross-section
will be slowed before
reaching the detector. 
This can be understood
by considering the evolution
of the velocity of a macro
as it passes through a medium

\begin{equation}
v(x) = v_{0} e^{-\langle \rho \Delta\rangle \frac{\sigma_x}{M_x}}\,,
\end{equation}
where $\langle \rho \Delta\rangle$ is
the integrated column density traversed 
defined as
\begin{equation}
\langle \rho \Delta\rangle = \int_l \rho(l) dx\,,
\end{equation}
where $l$ represents 
the trajectory of the macro, 
$v_0$ is
the initial velocity of the 
macro and $\frac{\sigma_x}{M_x}$
is the reduced cross-section.
Indeed, this is how the
upper bounds are generally
derived for the various exclusion
regions in Figure \ref{fig:exclusion}.

A similar scenario is expected
to manifest for macros with
a significant amount of charge.
If macros were to possess a large 
charge, they would have transferred 
a significant fraction of
their initial energy to the overlying
layers of rock or the outer
layer of a neutron star and thus be slowed
down before reaching the
detector.

Thus, we will require, 
\begin{equation}\label{delE}
\delta E \ll \frac{1}{2}M_x v_x^2\,,
\end{equation}
where the energy loss
is from both mechanisms
in \eqref{totaldedx}.
This criteria will
be used in determining
the upper bound on the charge 
constraints of 
a macro. These considerations
for ancient mica and the MACRO
experiment reveal that any 
upper bounds are 
similar to 
those derived by requiring
that $\epsilon \lesssim 10^{-5}$.

For objects of a fixed 
physical size (and hence geometric
cross-section), there is a 
maximum mass before the object
becomes a black hole. This
is illustrated
in the results figures as
the white hatched region
on the right. The high mass
boundary was determined
by solving for the mass
corresponding to the
Schwarzchild radius\cite{carroll}
\begin{equation}\label{schwarzchild}
M_{upper} = \sqrt{\frac{\sigma_x}{\pi}}
\frac{c^2}{2G}\,.
\end{equation}
Objects heavier than this mass
should not exist, as they would be
denser than black holes
of the same Schwarzchild radius.
These regions are hatched with $+$
symbols in the results figures.

\subsection*{Presentation
of results}
In this work, we have
introduced a third parameter
to describe a physical
attribute of a macro, $Q_x$.
Thus, the results should
be presented in a three dimensional
parameter space. However, this
will not be as informative
as in the two dimensional analog
when we considered only $\sigma_x$
and $M_x$. Instead we present,
as our results,
the two dimensional parameter
space of $Q_x$ and $M_x$ for 
slices of constant $\sigma_x$.
We hope that by presenting
results for several value of
$\sigma_x$, that the overall picture 
of
the constraints and their
evolution as we change
$\sigma_x$ becomes clear
to the reader.

For a given $\sigma_x$, there
exists constraints for
some range of $M_x$ independent of
the charge $Q_x$ of the macro. 
Thus, some values of
$Q_x$ are constrained in
abundance already by
the elastic scattering 
considerations. This results
in regions constrained for some
range of masses upto
some value of $Q_x$ in Figures
\ref{fig:money0} - \ref{fig:money5},
corresponding to the minimum $Q_x$ values
required for a track to have been
left purely by Coulomb interactions.
These constraints from
elastic scattering are presented
in the same color as the constraints
from the charge of the macro but without
the diagonal hatching.
{  
We note that Figure \ref{fig:money0}
is at a sufficiently low elastic cross-section $\sigma_x$ that
it is a characteristic result for all smaller elastic cross sections, i.e.
these constraints apply to objects that are phenomenologically similar
to the charged Planck-scale relics considered in reference 
\cite{1910.01285}. If such objects existed and were not
electrically charged, they would be
particles with effectively no non-gravitational interaction with the
constituents of the standard model.
}

The lower bounds in $M_x$ come from
the requirement that the macro
not loose a significant fraction
of its energy before reaching the 
appropriate depth in either the Earth
(as in the case of the ancient mica
or the MACRO detector) or a
neutron star.
Thus,
for the constraints from large scale
structure, there is no lower bound
on $M_x$ for bounds
from $\sigma_{elastic}$ independent
of Coulomb interactions.

For the constraints
derived using MACRO, mica
and 4U 1820-30, a sufficiently
large $\sigma_x$ results in
a minimum mass on the constraints
due to the upper bound
from \eqref{delE} with
the energy loss
in this case dominated by
elastic scattering.

We find that charges up to 
$Q_x \sim 10^6$ are constrained
using the MACRO and mica null
results. This is significantly
above any values
of Z that exist in the periodic 
table. However, it
is reasonable to expect
that larger values of $Q_x$
would deposit more energy.
Thus, one might expect
that larger values of $Q_x$
than those constrained here
would also be ruled out
based on the null observation
of tracks in MACRO and mica.
However, as our phenomenological
model breaks down around
$Q_x \sim 10^6$,
we stop placing
constraints at these $Q_x$ 
values, even though it is 
likely constraints exist at 
larger values of $Q_x$.
More conservatively, we
expect the results to hold upto
an atomic number of order
$\mathcal{O}(100-1000)$;
however, we present the entire 
range of constraints.

The neutron star
constraints are at higher values
of $Q_x$, mainly due to
the higher linear energy
deposition threshold value
required to trigger thermonuclear
runaway.

The results from large scale
structure prohibit large values
of $Q_x/M_x$. We plot
both the results from
self-interaction
and ISM analysis because the CMB 
bound is subject to additional
scrutiny. A theory describing
the formation of macros 
in the early Universe is required
for this bound to be
taken at the same
level of rigor as the other
late Universe constraints.

We note that
we are constraining objects with 
physical sizes below
that which are normally
associated with 
macroscopic dark matter
\cite{jacobs2015macro}, i.e.
objects smaller than about the 
size of a nuclei.

Finally, we also note that
our results constrain
all charge values
between the maximum and minimum
limits
and not only charge values
$Q_x = ne$ or $Q_x = \frac{n}{3}e$,
where $n$ is an integer.

\section{Conclusion}
We have produced constraints on the
maximum charge of macros
from {  phenomenological considerations} on a
variety of scales. We have
used galactic dynamics,
CMB measurements and galaxy
cluster considerations to constrain
dark matter charge 
constraints on large scales.
On terrestrial scales, the
lack of any tracks
observed in an slab of mica
exposed to the bombardment 
exposed over geologic timescales,
and in the MACRO
experiment were also used to 
constrain small mass macros.
Finally, the duration
between back-to-back superbursts
on 4U 1820-30 was used
to constrain intermediate mass
macros. 

{  In Figures \ref{fig:money0}-\ref{fig:money5},
we have shown the regions of parameter space where
charged macro candidates cannot contribute all the dark matter
that is observed on a variety of cosmological scales.}

It is of particular interest
to note that the results from 
MACRO seem to exclude
macros over the appropriate mass range
and geometric cross-sections
from 
being charged at all, as
these results exclude charges
down to approximately $\frac{e}{3}$,
which is the smallest known
quantized charged value, assuming
macros to me made of standard model
particles. 

It is interesting to return
to the question of binding
energies of macros first raised
in Section I. Considering the results
derived, and the line
representing objects
with iron peak elements
binding energy in the
result figures, we find that
one would not expect
objects with large $Q_x$ values 
and small masses to be bound.
However,
the primary
concern in this work
has been to constrain the
allowed charge values for macros
to contribute all of the
dark matter, based
on purely observational grounds as
there currently exists no
concrete theory describing
the formation of a macro
and its subsequent binding energy.

\section{Acknowledgements}
This work was partially supported by Department of
Energy grant de-sc0009946 to the particle astrophysics
theory group at CWRU.
The author would like to thank Alexis Plascencia,
Pavel 
Fileviez Perez and
Glenn Starkman 
for initial
discussions on this manuscript.

\bibliographystyle{apsrev4-1}
\bibliography{Charge}

\begin{thebibliography}{52}%
\makeatletter
\providecommand \@ifxundefined [1]{%
 \@ifx{#1\undefined}
}%
\providecommand \@ifnum [1]{%
 \ifnum #1\expandafter \@firstoftwo
 \else \expandafter \@secondoftwo
 \fi
}%
\providecommand \@ifx [1]{%
 \ifx #1\expandafter \@firstoftwo
 \else \expandafter \@secondoftwo
 \fi
}%
\providecommand \natexlab [1]{#1}%
\providecommand \enquote  [1]{``#1''}%
\providecommand \bibnamefont  [1]{#1}%
\providecommand \bibfnamefont [1]{#1}%
\providecommand \citenamefont [1]{#1}%
\providecommand \href@noop [0]{\@secondoftwo}%
\providecommand \href [0]{\begingroup \@sanitize@url \@href}%
\providecommand \@href[1]{\@@startlink{#1}\@@href}%
\providecommand \@@href[1]{\endgroup#1\@@endlink}%
\providecommand \@sanitize@url [0]{\catcode `\\12\catcode `\$12\catcode
  `\&12\catcode `\#12\catcode `\^12\catcode `\_12\catcode `\%12\relax}%
\providecommand \@@startlink[1]{}%
\providecommand \@@endlink[0]{}%
\providecommand \url  [0]{\begingroup\@sanitize@url \@url }%
\providecommand \@url [1]{\endgroup\@href {#1}{\urlprefix }}%
\providecommand \urlprefix  [0]{URL }%
\providecommand \Eprint [0]{\href }%
\providecommand \doibase [0]{http://dx.doi.org/}%
\providecommand \selectlanguage [0]{\@gobble}%
\providecommand \bibinfo  [0]{\@secondoftwo}%
\providecommand \bibfield  [0]{\@secondoftwo}%
\providecommand \translation [1]{[#1]}%
\providecommand \BibitemOpen [0]{}%
\providecommand \bibitemStop [0]{}%
\providecommand \bibitemNoStop [0]{.\EOS\space}%
\providecommand \EOS [0]{\spacefactor3000\relax}%
\providecommand \BibitemShut  [1]{\csname bibitem#1\endcsname}%
\let\auto@bib@innerbib\@empty
\bibitem [{\citenamefont {Tanabashi}\ \emph {et~al.}(2018)\citenamefont
  {Tanabashi} \emph {et~al.}}]{PDG}%
  \BibitemOpen
  \bibfield  {author} {\bibinfo {author} {\bibfnamefont {M.}~\bibnamefont
  {Tanabashi}} \emph {et~al.} (\bibinfo {collaboration} {Particle Data
  Group}),\ }\href {\doibase 10.1103/PhysRevD.98.030001} {\bibfield  {journal}
  {\bibinfo  {journal} {Phys. Rev. D}\ }\textbf {\bibinfo {volume} {98}},\
  \bibinfo {pages} {030001} (\bibinfo {year} {2018})}\BibitemShut {NoStop}%
\bibitem [{\citenamefont {Peccei}\ and\ \citenamefont {Quinn}(1977)}]{axion}%
  \BibitemOpen
  \bibfield  {author} {\bibinfo {author} {\bibfnamefont {R.~D.}\ \bibnamefont
  {Peccei}}\ and\ \bibinfo {author} {\bibfnamefont {H.~R.}\ \bibnamefont
  {Quinn}},\ }\href {\doibase 10.1103/PhysRevLett.38.1440} {\bibfield
  {journal} {\bibinfo  {journal} {Phys. Rev. Lett.}\ }\textbf {\bibinfo
  {volume} {38}},\ \bibinfo {pages} {1440} (\bibinfo {year}
  {1977})}\BibitemShut {NoStop}%
\bibitem [{\citenamefont {Wilczek}(1978)}]{Wilczek_axion}%
  \BibitemOpen
  \bibfield  {author} {\bibinfo {author} {\bibfnamefont {F.}~\bibnamefont
  {Wilczek}},\ }\href {\doibase 10.1103/PhysRevLett.40.279} {\bibfield
  {journal} {\bibinfo  {journal} {Phys. Rev. Lett.}\ }\textbf {\bibinfo
  {volume} {40}},\ \bibinfo {pages} {279} (\bibinfo {year} {1978})}\BibitemShut
  {NoStop}%
\bibitem [{\citenamefont {Weinberg}(1978)}]{Weinberg_axion}%
  \BibitemOpen
  \bibfield  {author} {\bibinfo {author} {\bibfnamefont {S.}~\bibnamefont
  {Weinberg}},\ }\href {\doibase 10.1103/PhysRevLett.40.223} {\bibfield
  {journal} {\bibinfo  {journal} {Phys. Rev. Lett.}\ }\textbf {\bibinfo
  {volume} {40}},\ \bibinfo {pages} {223} (\bibinfo {year} {1978})}\BibitemShut
  {NoStop}%
\bibitem [{\citenamefont {Witten}(1984)}]{PhysRevD.30.272}%
  \BibitemOpen
  \bibfield  {author} {\bibinfo {author} {\bibfnamefont {E.}~\bibnamefont
  {Witten}},\ }\href {\doibase 10.1103/physrevd.30.272} {\bibfield  {journal}
  {\bibinfo  {journal} {Physical Review D}\ }\textbf {\bibinfo {volume} {30}},\
  \bibinfo {pages} {272} (\bibinfo {year} {1984})}\BibitemShut {NoStop}%
\bibitem [{\citenamefont {Lynn}\ \emph {et~al.}(1990)\citenamefont {Lynn},
  \citenamefont {Nelson},\ and\ \citenamefont {Tetradis}}]{LYNN1990186}%
  \BibitemOpen
  \bibfield  {author} {\bibinfo {author} {\bibfnamefont {B.~W.}\ \bibnamefont
  {Lynn}}, \bibinfo {author} {\bibfnamefont {A.~E.}\ \bibnamefont {Nelson}}, \
  and\ \bibinfo {author} {\bibfnamefont {N.}~\bibnamefont {Tetradis}},\ }\href
  {\doibase 10.1016/0550-3213(90)90614-j} {\bibfield  {journal} {\bibinfo
  {journal} {Nuclear Physics B}\ }\textbf {\bibinfo {volume} {345}},\ \bibinfo
  {pages} {186} (\bibinfo {year} {1990})}\BibitemShut {NoStop}%
\bibitem [{\citenamefont {Lynn}(2010)}]{1005.2124}%
  \BibitemOpen
  \bibfield  {author} {\bibinfo {author} {\bibfnamefont {B.~W.}\ \bibnamefont
  {Lynn}},\ }\href@noop {} {\enquote {\bibinfo {title} {Liquid phases in
  {SU(3)} {C}hiral {P}erturbation {T}heory: {D}rops of {S}trange {C}hiral
  {N}ucleon {L}iquid and {O}rdinary {C}hiral {H}eavy {N}uclear {L}iquid},}\ }
  (\bibinfo {year} {2010}),\ \Eprint {http://arxiv.org/abs/arXiv:1005.2124}
  {arXiv:1005.2124} \BibitemShut {NoStop}%
\bibitem [{\citenamefont {Zhitnitsky}(2003)}]{hep-ph/0202161}%
  \BibitemOpen
  \bibfield  {author} {\bibinfo {author} {\bibfnamefont {A.~R.}\ \bibnamefont
  {Zhitnitsky}},\ }\href {\doibase 10.1088/1475-7516/2003/10/010} {\bibfield
  {journal} {\bibinfo  {journal} {JCAP}\ }\textbf {\bibinfo {volume} {0310}},\
  \bibinfo {pages} {010} (\bibinfo {year} {2003})}\BibitemShut {NoStop}%
\bibitem [{\citenamefont {Pont{\'{o}}n}\ \emph {et~al.}(2019)\citenamefont
  {Pont{\'{o}}n}, \citenamefont {Bai},\ and\ \citenamefont {Jain}}]{Pontn2019}%
  \BibitemOpen
  \bibfield  {author} {\bibinfo {author} {\bibfnamefont {E.}~\bibnamefont
  {Pont{\'{o}}n}}, \bibinfo {author} {\bibfnamefont {Y.}~\bibnamefont {Bai}}, \
  and\ \bibinfo {author} {\bibfnamefont {B.}~\bibnamefont {Jain}},\ }\href
  {\doibase 10.1007/s13130-019-11194-5} {\bibfield  {journal} {\bibinfo
  {journal} {Journal of High Energy Physics}\ }\textbf {\bibinfo {volume}
  {2019}} (\bibinfo {year} {2019}),\ 10.1007/s13130-019-11194-5}\BibitemShut
  {NoStop}%
\bibitem [{\citenamefont {Jacobs}\ \emph
  {et~al.}(2015{\natexlab{a}})\citenamefont {Jacobs}, \citenamefont
  {Starkman},\ and\ \citenamefont {Lynn}}]{jacobs2015macro}%
  \BibitemOpen
  \bibfield  {author} {\bibinfo {author} {\bibfnamefont {D.~M.}\ \bibnamefont
  {Jacobs}}, \bibinfo {author} {\bibfnamefont {G.~D.}\ \bibnamefont
  {Starkman}}, \ and\ \bibinfo {author} {\bibfnamefont {B.~W.}\ \bibnamefont
  {Lynn}},\ }\href {\doibase 10.1093/mnras/stv774} {\bibfield  {journal}
  {\bibinfo  {journal} {Monthly Notices of the Royal Astronomical Society}\
  }\textbf {\bibinfo {volume} {450}},\ \bibinfo {pages} {3418} (\bibinfo {year}
  {2015}{\natexlab{a}})}\BibitemShut {NoStop}%
\bibitem [{\citenamefont {Jacobs}\ \emph
  {et~al.}(2015{\natexlab{b}})\citenamefont {Jacobs}, \citenamefont {Weltman},\
  and\ \citenamefont {Starkman}}]{jacobs2015resonant}%
  \BibitemOpen
  \bibfield  {author} {\bibinfo {author} {\bibfnamefont {D.~M.}\ \bibnamefont
  {Jacobs}}, \bibinfo {author} {\bibfnamefont {A.}~\bibnamefont {Weltman}}, \
  and\ \bibinfo {author} {\bibfnamefont {G.~D.}\ \bibnamefont {Starkman}},\
  }\href {\doibase 10.1103/physrevd.91.115023} {\bibfield  {journal} {\bibinfo
  {journal} {Physical Review D}\ }\textbf {\bibinfo {volume} {91}},\ \bibinfo
  {pages} {115023} (\bibinfo {year} {2015}{\natexlab{b}})}\BibitemShut
  {NoStop}%
\bibitem [{\citenamefont {Sidhu}\ \emph
  {et~al.}(2019{\natexlab{a}})\citenamefont {Sidhu}, \citenamefont {Scherrer},\
  and\ \citenamefont {Starkman}}]{1907.06674}%
  \BibitemOpen
  \bibfield  {author} {\bibinfo {author} {\bibfnamefont {J.~S.}\ \bibnamefont
  {Sidhu}}, \bibinfo {author} {\bibfnamefont {R.~J.}\ \bibnamefont {Scherrer}},
  \ and\ \bibinfo {author} {\bibfnamefont {G.}~\bibnamefont {Starkman}},\
  }\href@noop {} {\enquote {\bibinfo {title} {Death by dark matter},}\ }
  (\bibinfo {year} {2019}{\natexlab{a}}),\ \Eprint
  {http://arxiv.org/abs/arXiv:1907.06674} {arXiv:1907.06674} \BibitemShut
  {NoStop}%
\bibitem [{\citenamefont {Sidhu}\ and\ \citenamefont
  {Starkman}(2019{\natexlab{a}})}]{1908.00557}%
  \BibitemOpen
  \bibfield  {author} {\bibinfo {author} {\bibfnamefont {J.~S.}\ \bibnamefont
  {Sidhu}}\ and\ \bibinfo {author} {\bibfnamefont {G.}~\bibnamefont
  {Starkman}},\ }\href@noop {} {\enquote {\bibinfo {title} {Macroscopic dark
  matter constraints from bolide camera networks},}\ } (\bibinfo {year}
  {2019}{\natexlab{a}}),\ \Eprint {http://arxiv.org/abs/arXiv:1908.00557}
  {arXiv:1908.00557} \BibitemShut {NoStop}%
\bibitem [{\citenamefont {Sidhu}\ \emph
  {et~al.}(2019{\natexlab{b}})\citenamefont {Sidhu}, \citenamefont {Abraham},
  \citenamefont {Covault},\ and\ \citenamefont {Starkman}}]{Sidhu:2018auv}%
  \BibitemOpen
  \bibfield  {author} {\bibinfo {author} {\bibfnamefont {J.~S.}\ \bibnamefont
  {Sidhu}}, \bibinfo {author} {\bibfnamefont {R.~M.}\ \bibnamefont {Abraham}},
  \bibinfo {author} {\bibfnamefont {C.}~\bibnamefont {Covault}}, \ and\
  \bibinfo {author} {\bibfnamefont {G.}~\bibnamefont {Starkman}},\ }\href
  {\doibase 10.1088/1475-7516/2019/02/037} {\bibfield  {journal} {\bibinfo
  {journal} {JCAP}\ }\textbf {\bibinfo {volume} {1902}},\ \bibinfo {pages}
  {037} (\bibinfo {year} {2019}{\natexlab{b}})},\ \Eprint
  {http://arxiv.org/abs/1808.06978} {arXiv:1808.06978 [astro-ph.HE]}
  \BibitemShut {NoStop}%
\bibitem [{\citenamefont {Sidhu}\ \emph
  {et~al.}(2019{\natexlab{c}})\citenamefont {Sidhu}, \citenamefont {Starkman},\
  and\ \citenamefont {Harvey}}]{1905.10025}%
  \BibitemOpen
  \bibfield  {author} {\bibinfo {author} {\bibfnamefont {J.~S.}\ \bibnamefont
  {Sidhu}}, \bibinfo {author} {\bibfnamefont {G.}~\bibnamefont {Starkman}}, \
  and\ \bibinfo {author} {\bibfnamefont {R.}~\bibnamefont {Harvey}},\
  }\href@noop {} {\bibfield  {journal} {\bibinfo  {journal} {Physical Review
  D}\ }\textbf {\bibinfo {volume} {100}} (\bibinfo {year}
  {2019}{\natexlab{c}})}\BibitemShut {NoStop}%
\bibitem [{\citenamefont {De~Rujula}\ and\ \citenamefont
  {Glashow}(1984)}]{DeRujula:1984axn}%
  \BibitemOpen
  \bibfield  {author} {\bibinfo {author} {\bibfnamefont {A.}~\bibnamefont
  {De~Rujula}}\ and\ \bibinfo {author} {\bibfnamefont {S.~L.}\ \bibnamefont
  {Glashow}},\ }\href {\doibase 10.1038/312734a0} {\bibfield  {journal}
  {\bibinfo  {journal} {Nature}\ }\textbf {\bibinfo {volume} {312}},\ \bibinfo
  {pages} {734} (\bibinfo {year} {1984})}\BibitemShut {NoStop}%
\bibitem [{\citenamefont {Price}(1988)}]{Price:1988ge}%
  \BibitemOpen
  \bibfield  {author} {\bibinfo {author} {\bibfnamefont {P.~B.}\ \bibnamefont
  {Price}},\ }\href {\doibase 10.1103/PhysRevD.38.3813} {\bibfield  {journal}
  {\bibinfo  {journal} {Physical Review D}\ }\textbf {\bibinfo {volume} {38}},\
  \bibinfo {pages} {3813} (\bibinfo {year} {1988})}\BibitemShut {NoStop}%
\bibitem [{\citenamefont {Graham}\ \emph {et~al.}(2018)\citenamefont {Graham},
  \citenamefont {Janish}, \citenamefont {Narayan}, \citenamefont {Rajendran},\
  and\ \citenamefont {Riggins}}]{1805.07381}%
  \BibitemOpen
  \bibfield  {author} {\bibinfo {author} {\bibfnamefont {P.~W.}\ \bibnamefont
  {Graham}}, \bibinfo {author} {\bibfnamefont {R.}~\bibnamefont {Janish}},
  \bibinfo {author} {\bibfnamefont {V.}~\bibnamefont {Narayan}}, \bibinfo
  {author} {\bibfnamefont {S.}~\bibnamefont {Rajendran}}, \ and\ \bibinfo
  {author} {\bibfnamefont {P.}~\bibnamefont {Riggins}},\ }\href {\doibase
  10.1103/physrevd.98.115027} {\bibfield  {journal} {\bibinfo  {journal}
  {Physical Review D}\ }\textbf {\bibinfo {volume} {98}},\ \bibinfo {pages}
  {115027} (\bibinfo {year} {2018})}\BibitemShut {NoStop}%
\bibitem [{\citenamefont {et~al. and}(2002)}]{hepex0207020}%
  \BibitemOpen
  \bibfield  {author} {\bibinfo {author} {\bibfnamefont {M.~A.}\ \bibnamefont
  {et~al. and}},\ }\href {\doibase 10.1140/epjc/s2002-01046-9} {\bibfield
  {journal} {\bibinfo  {journal} {The European Physical Journal C}\ }\textbf
  {\bibinfo {volume} {25}},\ \bibinfo {pages} {511} (\bibinfo {year}
  {2002})}\BibitemShut {NoStop}%
\bibitem [{\citenamefont {Alcock}\ \emph {et~al.}(2001)\citenamefont {Alcock}
  \emph {et~al.}}]{Alcock2001}%
  \BibitemOpen
  \bibfield  {author} {\bibinfo {author} {\bibfnamefont {C.}~\bibnamefont
  {Alcock}} \emph {et~al.},\ }\href {\doibase 10.1086/319636} {\bibfield
  {journal} {\bibinfo  {journal} {The Astrophysical Journal}\ }\textbf
  {\bibinfo {volume} {550}},\ \bibinfo {pages} {L169} (\bibinfo {year}
  {2001})}\BibitemShut {NoStop}%
\bibitem [{\citenamefont {Tisserand}\ \emph {et~al.}(2007)\citenamefont
  {Tisserand} \emph {et~al.}}]{astro-ph/0607207}%
  \BibitemOpen
  \bibfield  {author} {\bibinfo {author} {\bibfnamefont {P.}~\bibnamefont
  {Tisserand}} \emph {et~al.},\ }\href {\doibase 10.1051/0004-6361:20066017}
  {\bibfield  {journal} {\bibinfo  {journal} {Astronomy {\&} Astrophysics}\
  }\textbf {\bibinfo {volume} {469}},\ \bibinfo {pages} {387} (\bibinfo {year}
  {2007})}\BibitemShut {NoStop}%
\bibitem [{\citenamefont {Carr}\ \emph {et~al.}(2010)\citenamefont {Carr},
  \citenamefont {Kohri}, \citenamefont {Sendouda},\ and\ \citenamefont
  {Yokoyama}}]{0912.5297}%
  \BibitemOpen
  \bibfield  {author} {\bibinfo {author} {\bibfnamefont {B.~J.}\ \bibnamefont
  {Carr}}, \bibinfo {author} {\bibfnamefont {K.}~\bibnamefont {Kohri}},
  \bibinfo {author} {\bibfnamefont {Y.}~\bibnamefont {Sendouda}}, \ and\
  \bibinfo {author} {\bibfnamefont {J.}~\bibnamefont {Yokoyama}},\ }\href
  {\doibase 10.1103/physrevd.81.104019} {\bibfield  {journal} {\bibinfo
  {journal} {Physical Review D}\ }\textbf {\bibinfo {volume} {81}},\ \bibinfo
  {pages} {104019} (\bibinfo {year} {2010})}\BibitemShut {NoStop}%
\bibitem [{\citenamefont {Griest}\ \emph {et~al.}(2013)\citenamefont {Griest},
  \citenamefont {Cieplak},\ and\ \citenamefont {Lehner}}]{Griest2013}%
  \BibitemOpen
  \bibfield  {author} {\bibinfo {author} {\bibfnamefont {K.}~\bibnamefont
  {Griest}}, \bibinfo {author} {\bibfnamefont {A.~M.}\ \bibnamefont {Cieplak}},
  \ and\ \bibinfo {author} {\bibfnamefont {M.~J.}\ \bibnamefont {Lehner}},\
  }\href {\doibase 10.1103/physrevlett.111.181302} {\bibfield  {journal}
  {\bibinfo  {journal} {Physical Review Letters}\ }\textbf {\bibinfo {volume}
  {111}},\ \bibinfo {pages} {181302} (\bibinfo {year} {2013})}\BibitemShut
  {NoStop}%
\bibitem [{\citenamefont {Niikura}\ \emph {et~al.}(2019)\citenamefont {Niikura}
  \emph {et~al.}}]{Niikura2019}%
  \BibitemOpen
  \bibfield  {author} {\bibinfo {author} {\bibfnamefont {H.}~\bibnamefont
  {Niikura}} \emph {et~al.},\ }\href {\doibase 10.1038/s41550-019-0723-1}
  {\bibfield  {journal} {\bibinfo  {journal} {Nature Astronomy}\ }\textbf
  {\bibinfo {volume} {3}},\ \bibinfo {pages} {524} (\bibinfo {year}
  {2019})}\BibitemShut {NoStop}%
\bibitem [{\citenamefont {Smyth}\ \emph {et~al.}(2019)\citenamefont {Smyth},
  \citenamefont {Profumo}, \citenamefont {English}, \citenamefont {Jeltema},
  \citenamefont {McKinnon},\ and\ \citenamefont {Guhathakurta}}]{1910.01285}%
  \BibitemOpen
  \bibfield  {author} {\bibinfo {author} {\bibfnamefont {N.}~\bibnamefont
  {Smyth}}, \bibinfo {author} {\bibfnamefont {S.}~\bibnamefont {Profumo}},
  \bibinfo {author} {\bibfnamefont {S.}~\bibnamefont {English}}, \bibinfo
  {author} {\bibfnamefont {T.}~\bibnamefont {Jeltema}}, \bibinfo {author}
  {\bibfnamefont {K.}~\bibnamefont {McKinnon}}, \ and\ \bibinfo {author}
  {\bibfnamefont {P.}~\bibnamefont {Guhathakurta}},\ }\href@noop {} {\enquote
  {\bibinfo {title} {Updated constraints on asteroid-mass primordial black
  holes as dark matter},}\ } (\bibinfo {year} {2019}),\ \Eprint
  {http://arxiv.org/abs/arXiv:1910.01285} {arXiv:1910.01285} \BibitemShut
  {NoStop}%
\bibitem [{\citenamefont {Wilkinson}\ \emph {et~al.}(2013)\citenamefont
  {Wilkinson}, \citenamefont {Lesgourgues},\ and\ \citenamefont
  {Boehm}}]{1309.7588}%
  \BibitemOpen
  \bibfield  {author} {\bibinfo {author} {\bibfnamefont {R.~J.}\ \bibnamefont
  {Wilkinson}}, \bibinfo {author} {\bibfnamefont {J.}~\bibnamefont
  {Lesgourgues}}, \ and\ \bibinfo {author} {\bibfnamefont {C.}~\bibnamefont
  {Boehm}},\ }\href {\doibase 10.1088/1475-7516/2014/04/026} {\bibfield
  {journal} {\bibinfo  {journal} {JCAP}\ }\textbf {\bibinfo {volume} {4}},\
  \bibinfo {pages} {026} (\bibinfo {year} {2013})},\ \Eprint
  {http://arxiv.org/abs/arXiv:1309.7588} {arXiv:1309.7588} \BibitemShut
  {NoStop}%
\bibitem [{\citenamefont {B{\oe}hm}\ \emph {et~al.}(2001)\citenamefont
  {B{\oe}hm}, \citenamefont {Fayet},\ and\ \citenamefont {Schaeffer}}]{Bhm}%
  \BibitemOpen
  \bibfield  {author} {\bibinfo {author} {\bibfnamefont {C.}~\bibnamefont
  {B{\oe}hm}}, \bibinfo {author} {\bibfnamefont {P.}~\bibnamefont {Fayet}}, \
  and\ \bibinfo {author} {\bibfnamefont {R.}~\bibnamefont {Schaeffer}},\ }\href
  {\doibase 10.1016/s0370-2693(01)01060-7} {\bibfield  {journal} {\bibinfo
  {journal} {Physics Letters B}\ }\textbf {\bibinfo {volume} {518}},\ \bibinfo
  {pages} {8} (\bibinfo {year} {2001})}\BibitemShut {NoStop}%
\bibitem [{\citenamefont {Sidhu}\ and\ \citenamefont
  {Starkman}(2019{\natexlab{b}})}]{1912.04053}%
  \BibitemOpen
  \bibfield  {author} {\bibinfo {author} {\bibfnamefont {J.~S.}\ \bibnamefont
  {Sidhu}}\ and\ \bibinfo {author} {\bibfnamefont {G.~D.}\ \bibnamefont
  {Starkman}},\ }\href@noop {} {\enquote {\bibinfo {title} {Reconsidering
  astrophysical constraints on macroscopic dark matter},}\ } (\bibinfo {year}
  {2019}{\natexlab{b}}),\ \Eprint {http://arxiv.org/abs/arXiv:1912.04053}
  {arXiv:1912.04053} \BibitemShut {NoStop}%
\bibitem [{Note1()}]{Note1}%
  \BibitemOpen
  \bibinfo {note} {This is the distribution of macro velocities in a
  non-orbiting frame moving with the Galaxy. When considering the velocity of
  macros impacting the atmosphere, \protect \textup {\hbox {\mathsurround \z@
  \protect \normalfont (\ignorespaces \ref {maxwellian}\unskip \@@italiccorr
  )}} is modified by the motion of the Sun and Earth in that frame, and by the
  Sun's and Earth's gravitational potential. We have taken into account these
  effects (as explained, for example, in \cite {Freese2013}), except the
  negligible effect of Earth's gravitational potential.}\BibitemShut {Stop}%
\bibitem [{\citenamefont {Davidson}\ \emph {et~al.}(2000)\citenamefont
  {Davidson}, \citenamefont {Hannestad},\ and\ \citenamefont
  {Raffelt}}]{Davidson2000}%
  \BibitemOpen
  \bibfield  {author} {\bibinfo {author} {\bibfnamefont {S.}~\bibnamefont
  {Davidson}}, \bibinfo {author} {\bibfnamefont {S.}~\bibnamefont {Hannestad}},
  \ and\ \bibinfo {author} {\bibfnamefont {G.}~\bibnamefont {Raffelt}},\ }\href
  {\doibase 10.1088/1126-6708/2000/05/003} {\bibfield  {journal} {\bibinfo
  {journal} {Journal of High Energy Physics}\ }\textbf {\bibinfo {volume}
  {2000}},\ \bibinfo {pages} {003} (\bibinfo {year} {2000})}\BibitemShut
  {NoStop}%
\bibitem [{\citenamefont {Dimopoulos}\ \emph {et~al.}(1990)\citenamefont
  {Dimopoulos}, \citenamefont {Eichler}, \citenamefont {Esmailzadeh},\ and\
  \citenamefont {Starkman}}]{Dimopoulos1990}%
  \BibitemOpen
  \bibfield  {author} {\bibinfo {author} {\bibfnamefont {S.}~\bibnamefont
  {Dimopoulos}}, \bibinfo {author} {\bibfnamefont {D.}~\bibnamefont {Eichler}},
  \bibinfo {author} {\bibfnamefont {R.}~\bibnamefont {Esmailzadeh}}, \ and\
  \bibinfo {author} {\bibfnamefont {G.~D.}\ \bibnamefont {Starkman}},\ }\href
  {\doibase 10.1103/physrevd.41.2388} {\bibfield  {journal} {\bibinfo
  {journal} {Physical Review D}\ }\textbf {\bibinfo {volume} {41}},\ \bibinfo
  {pages} {2388} (\bibinfo {year} {1990})}\BibitemShut {NoStop}%
\bibitem [{\citenamefont {Prinz}\ \emph {et~al.}(1998)\citenamefont {Prinz}
  \emph {et~al.}}]{Prinz1998}%
  \BibitemOpen
  \bibfield  {author} {\bibinfo {author} {\bibfnamefont {A.~A.}\ \bibnamefont
  {Prinz}} \emph {et~al.},\ }\href {\doibase 10.1103/physrevlett.81.1175}
  {\bibfield  {journal} {\bibinfo  {journal} {Physical Review Letters}\
  }\textbf {\bibinfo {volume} {81}},\ \bibinfo {pages} {1175} (\bibinfo {year}
  {1998})}\BibitemShut {NoStop}%
\bibitem [{\citenamefont {Chang}\ \emph {et~al.}(2018)\citenamefont {Chang},
  \citenamefont {Essig},\ and\ \citenamefont {McDermott}}]{Chang2018}%
  \BibitemOpen
  \bibfield  {author} {\bibinfo {author} {\bibfnamefont {J.~H.}\ \bibnamefont
  {Chang}}, \bibinfo {author} {\bibfnamefont {R.}~\bibnamefont {Essig}}, \ and\
  \bibinfo {author} {\bibfnamefont {S.~D.}\ \bibnamefont {McDermott}},\ }\href
  {\doibase 10.1007/jhep09(2018)051} {\bibfield  {journal} {\bibinfo  {journal}
  {Journal of High Energy Physics}\ }\textbf {\bibinfo {volume} {2018}}
  (\bibinfo {year} {2018}),\ 10.1007/jhep09(2018)051}\BibitemShut {NoStop}%
\bibitem [{\citenamefont {Lehmann}\ \emph {et~al.}(2019)\citenamefont
  {Lehmann}, \citenamefont {Johnson}, \citenamefont {Profumo},\ and\
  \citenamefont {Schwemberger}}]{1906.06348}%
  \BibitemOpen
  \bibfield  {author} {\bibinfo {author} {\bibfnamefont {B.~V.}\ \bibnamefont
  {Lehmann}}, \bibinfo {author} {\bibfnamefont {C.}~\bibnamefont {Johnson}},
  \bibinfo {author} {\bibfnamefont {S.}~\bibnamefont {Profumo}}, \ and\
  \bibinfo {author} {\bibfnamefont {T.}~\bibnamefont {Schwemberger}},\ }\href
  {\doibase 10.1088/1475-7516/2019/10/046} {\bibfield  {journal} {\bibinfo
  {journal} {Journal of Cosmology and Astroparticle Physics}\ }\textbf
  {\bibinfo {volume} {2019}},\ \bibinfo {pages} {046} (\bibinfo {year}
  {2019})}\BibitemShut {NoStop}%
\bibitem [{\citenamefont {Bai}\ and\ \citenamefont
  {Berger}(2019)}]{1912.02813}%
  \BibitemOpen
  \bibfield  {author} {\bibinfo {author} {\bibfnamefont {Y.}~\bibnamefont
  {Bai}}\ and\ \bibinfo {author} {\bibfnamefont {J.}~\bibnamefont {Berger}},\
  }\href@noop {} {\enquote {\bibinfo {title} {Nucleus capture by macroscopic
  dark matter},}\ } (\bibinfo {year} {2019}),\ \Eprint
  {http://arxiv.org/abs/arXiv:1912.02813} {arXiv:1912.02813} \BibitemShut
  {NoStop}%
\bibitem [{\citenamefont {Randall}\ \emph {et~al.}(2008)\citenamefont
  {Randall}, \citenamefont {Markevitch}, \citenamefont {Clowe}, \citenamefont
  {Gonzalez},\ and\ \citenamefont {Brada{\v{c}}}}]{Randall2008}%
  \BibitemOpen
  \bibfield  {author} {\bibinfo {author} {\bibfnamefont {S.~W.}\ \bibnamefont
  {Randall}}, \bibinfo {author} {\bibfnamefont {M.}~\bibnamefont {Markevitch}},
  \bibinfo {author} {\bibfnamefont {D.}~\bibnamefont {Clowe}}, \bibinfo
  {author} {\bibfnamefont {A.~H.}\ \bibnamefont {Gonzalez}}, \ and\ \bibinfo
  {author} {\bibfnamefont {M.}~\bibnamefont {Brada{\v{c}}}},\ }\href {\doibase
  10.1086/587859} {\bibfield  {journal} {\bibinfo  {journal} {The Astrophysical
  Journal}\ }\textbf {\bibinfo {volume} {679}},\ \bibinfo {pages} {1173}
  (\bibinfo {year} {2008})}\BibitemShut {NoStop}%
\bibitem [{\citenamefont {Chandrasekhar}(1943)}]{Chandrasekhar1943}%
  \BibitemOpen
  \bibfield  {author} {\bibinfo {author} {\bibfnamefont {S.}~\bibnamefont
  {Chandrasekhar}},\ }\href {\doibase 10.1086/144517} {\bibfield  {journal}
  {\bibinfo  {journal} {The Astrophysical Journal}\ }\textbf {\bibinfo {volume}
  {97}},\ \bibinfo {pages} {255} (\bibinfo {year} {1943})}\BibitemShut
  {NoStop}%
\bibitem [{\citenamefont {{Binney}}\ and\ \citenamefont
  {{Tremaine}}(2008)}]{galdyn}%
  \BibitemOpen
  \bibfield  {author} {\bibinfo {author} {\bibfnamefont {J.}~\bibnamefont
  {{Binney}}}\ and\ \bibinfo {author} {\bibfnamefont {S.}~\bibnamefont
  {{Tremaine}}},\ }\href@noop {} {\emph {\bibinfo {title} {Galactic Dynamics:
  Second Edition, by James Binney and Scott Tremaine.~ISBN 978-0-691-13026-2
  (HB).~Published by Princeton University Press, Princeton, NJ USA, 2008.}}}\
  (\bibinfo  {publisher} {Princeton University Press},\ \bibinfo {year}
  {2008})\BibitemShut {NoStop}%
\bibitem [{\citenamefont {Dolgov}\ \emph {et~al.}(2013)\citenamefont {Dolgov},
  \citenamefont {Dubovsky}, \citenamefont {Rubtsov},\ and\ \citenamefont
  {Tkachev}}]{PhysRevD.88.117701}%
  \BibitemOpen
  \bibfield  {author} {\bibinfo {author} {\bibfnamefont {A.~D.}\ \bibnamefont
  {Dolgov}}, \bibinfo {author} {\bibfnamefont {S.~L.}\ \bibnamefont
  {Dubovsky}}, \bibinfo {author} {\bibfnamefont {G.~I.}\ \bibnamefont
  {Rubtsov}}, \ and\ \bibinfo {author} {\bibfnamefont {I.~I.}\ \bibnamefont
  {Tkachev}},\ }\href {\doibase 10.1103/PhysRevD.88.117701} {\bibfield
  {journal} {\bibinfo  {journal} {Phys. Rev. D}\ }\textbf {\bibinfo {volume}
  {88}},\ \bibinfo {pages} {117701} (\bibinfo {year} {2013})}\BibitemShut
  {NoStop}%
\bibitem [{\citenamefont {Kadota}\ \emph {et~al.}(2016)\citenamefont {Kadota},
  \citenamefont {Sekiguchia},\ and\ \citenamefont {Tashirob}}]{1602.04009}%
  \BibitemOpen
  \bibfield  {author} {\bibinfo {author} {\bibfnamefont {K.}~\bibnamefont
  {Kadota}}, \bibinfo {author} {\bibfnamefont {T.}~\bibnamefont {Sekiguchia}},
  \ and\ \bibinfo {author} {\bibfnamefont {H.}~\bibnamefont {Tashirob}},\
  }\href@noop {} {\emph {\bibinfo {title} {A new constraint on millicharged
  dark matter from galaxy clusters}}},\ \bibinfo {type} {Tech. Rep.}\ \bibinfo
  {number} {CTPU-16-04}\ (\bibinfo {year} {2016})\BibitemShut {NoStop}%
\bibitem [{\citenamefont {et~al}(2008)}]{SRIM}%
  \BibitemOpen
  \bibfield  {author} {\bibinfo {author} {\bibfnamefont {Z.~J.}\ \bibnamefont
  {et~al}},\ }\href
  {https://www.amazon.com/SRIM-Stopping-Range-Ions-Matter/dp/B001QLHOTC?SubscriptionId=AKIAIOBINVZYXZQZ2U3A&tag=chimbori05-20&linkCode=xm2&camp=2025&creative=165953&creativeASIN=B001QLHOTC}
  {\emph {\bibinfo {title} {SRIM - The Stopping and Range of Ions in Matter}}}\
  (\bibinfo {year} {2008})\BibitemShut {NoStop}%
\bibitem [{\citenamefont {Spergel}\ and\ \citenamefont
  {Steinhardt}(2000)}]{PhysRevLett.84.3760}%
  \BibitemOpen
  \bibfield  {author} {\bibinfo {author} {\bibfnamefont {D.~N.}\ \bibnamefont
  {Spergel}}\ and\ \bibinfo {author} {\bibfnamefont {P.~J.}\ \bibnamefont
  {Steinhardt}},\ }\href {\doibase 10.1103/PhysRevLett.84.3760} {\bibfield
  {journal} {\bibinfo  {journal} {Phys. Rev. Lett.}\ }\textbf {\bibinfo
  {volume} {84}},\ \bibinfo {pages} {3760} (\bibinfo {year}
  {2000})}\BibitemShut {NoStop}%
\bibitem [{\citenamefont {Rocha}\ \emph {et~al.}(2013)\citenamefont {Rocha},
  \citenamefont {Peter}, \citenamefont {Bullock}, \citenamefont {Kaplinghat},
  \citenamefont {Garrison-Kimmel}, \citenamefont {O{\~{n}}orbe},\ and\
  \citenamefont {Moustakas}}]{Rocha2013}%
  \BibitemOpen
  \bibfield  {author} {\bibinfo {author} {\bibfnamefont {M.}~\bibnamefont
  {Rocha}}, \bibinfo {author} {\bibfnamefont {A.~H.~G.}\ \bibnamefont {Peter}},
  \bibinfo {author} {\bibfnamefont {J.~S.}\ \bibnamefont {Bullock}}, \bibinfo
  {author} {\bibfnamefont {M.}~\bibnamefont {Kaplinghat}}, \bibinfo {author}
  {\bibfnamefont {S.}~\bibnamefont {Garrison-Kimmel}}, \bibinfo {author}
  {\bibfnamefont {J.}~\bibnamefont {O{\~{n}}orbe}}, \ and\ \bibinfo {author}
  {\bibfnamefont {L.~A.}\ \bibnamefont {Moustakas}},\ }\href {\doibase
  10.1093/mnras/sts514} {\bibfield  {journal} {\bibinfo  {journal} {Monthly
  Notices of the Royal Astronomical Society}\ }\textbf {\bibinfo {volume}
  {430}},\ \bibinfo {pages} {81} (\bibinfo {year} {2013})}\BibitemShut
  {NoStop}%
\bibitem [{\citenamefont {Epstein}(1924)}]{Epstein}%
  \BibitemOpen
  \bibfield  {author} {\bibinfo {author} {\bibfnamefont {P.~S.}\ \bibnamefont
  {Epstein}},\ }\href {\doibase 10.1103/PhysRev.23.710} {\bibfield  {journal}
  {\bibinfo  {journal} {Phys. Rev.}\ }\textbf {\bibinfo {volume} {23}},\
  \bibinfo {pages} {710} (\bibinfo {year} {1924})}\BibitemShut {NoStop}%
\bibitem [{\citenamefont {Benito}\ \emph {et~al.}(2019)\citenamefont {Benito},
  \citenamefont {Cuoco},\ and\ \citenamefont {Iocco}}]{Benito2019}%
  \BibitemOpen
  \bibfield  {author} {\bibinfo {author} {\bibfnamefont {M.}~\bibnamefont
  {Benito}}, \bibinfo {author} {\bibfnamefont {A.}~\bibnamefont {Cuoco}}, \
  and\ \bibinfo {author} {\bibfnamefont {F.}~\bibnamefont {Iocco}},\ }\href
  {\doibase 10.1088/1475-7516/2019/03/033} {\bibfield  {journal} {\bibinfo
  {journal} {Journal of Cosmology and Astroparticle Physics}\ }\textbf
  {\bibinfo {volume} {2019}},\ \bibinfo {pages} {033} (\bibinfo {year}
  {2019})}\BibitemShut {NoStop}%
\bibitem [{\citenamefont {Bengaly}\ \emph {et~al.}(2019)\citenamefont
  {Bengaly}, \citenamefont {Clarkson},\ and\ \citenamefont
  {Maartens}}]{1908.04619}%
  \BibitemOpen
  \bibfield  {author} {\bibinfo {author} {\bibfnamefont {C.~A.~P.}\
  \bibnamefont {Bengaly}}, \bibinfo {author} {\bibfnamefont {C.}~\bibnamefont
  {Clarkson}}, \ and\ \bibinfo {author} {\bibfnamefont {R.}~\bibnamefont
  {Maartens}},\ }\href@noop {} {\enquote {\bibinfo {title} {The hubble constant
  tension with next generation galaxy surveys},}\ } (\bibinfo {year} {2019}),\
  \Eprint {http://arxiv.org/abs/arXiv:1908.04619} {arXiv:1908.04619}
  \BibitemShut {NoStop}%
\bibitem [{\citenamefont {Hellborg}(2011)}]{hellborg_2011}%
  \BibitemOpen
  \bibfield  {author} {\bibinfo {author} {\bibfnamefont {R.}~\bibnamefont
  {Hellborg}},\ }\href@noop {} {\emph {\bibinfo {title} {Electrostatic
  accelerators: fundamentals and applications}}}\ (\bibinfo  {publisher}
  {Springer},\ \bibinfo {year} {2011})\BibitemShut {NoStop}%
\bibitem [{978(2010)}]{9789048145102}%
  \BibitemOpen
  \href
  {https://www.amazon.com/Application-Particle-Materials-Technology-Science/dp/9048145104?SubscriptionId=AKIAIOBINVZYXZQZ2U3A&tag=chimbori05-20&linkCode=xm2&camp=2025&creative=165953&creativeASIN=9048145104}
  {\emph {\bibinfo {title} {Application of Particle and Laser Beams in
  Materials Technology (Nato Science Series E:)}}}\ (\bibinfo  {publisher}
  {Springer},\ \bibinfo {year} {2010})\BibitemShut {NoStop}%
\bibitem [{\citenamefont {Graham}\ \emph {et~al.}(2015)\citenamefont {Graham},
  \citenamefont {Rajendran},\ and\ \citenamefont {Varela}}]{1505.04444}%
  \BibitemOpen
  \bibfield  {author} {\bibinfo {author} {\bibfnamefont {P.~W.}\ \bibnamefont
  {Graham}}, \bibinfo {author} {\bibfnamefont {S.}~\bibnamefont {Rajendran}}, \
  and\ \bibinfo {author} {\bibfnamefont {J.}~\bibnamefont {Varela}},\ }\href
  {\doibase 10.1103/physrevd.92.063007} {\bibfield  {journal} {\bibinfo
  {journal} {Physical Review D}\ }\textbf {\bibinfo {volume} {92}},\ \bibinfo
  {pages} {063007} (\bibinfo {year} {2015})}\BibitemShut {NoStop}%
\bibitem [{\citenamefont {Timmes}\ and\ \citenamefont
  {Woosley}(1992)}]{Timmes1992}%
  \BibitemOpen
  \bibfield  {author} {\bibinfo {author} {\bibfnamefont {F.~X.}\ \bibnamefont
  {Timmes}}\ and\ \bibinfo {author} {\bibfnamefont {S.~E.}\ \bibnamefont
  {Woosley}},\ }\href {\doibase 10.1086/171746} {\bibfield  {journal} {\bibinfo
   {journal} {The Astrophysical Journal}\ }\textbf {\bibinfo {volume} {396}},\
  \bibinfo {pages} {649} (\bibinfo {year} {1992})}\BibitemShut {NoStop}%
\bibitem [{car(1970)}]{carroll}%
  \BibitemOpen
  \href
  {https://www.amazon.com/Spacetime-Geometry-International-Introduction-Relativity/dp/1292026634?SubscriptionId=AKIAIOBINVZYXZQZ2U3A&tag=chimbori05-20&linkCode=xm2&camp=2025&creative=165953&creativeASIN=1292026634}
  {\emph {\bibinfo {title} {Spacetime and Geometry: An Introduction to General
  Relativity}}}\ (\bibinfo  {publisher} {Pearson International},\ \bibinfo
  {year} {1970})\BibitemShut {NoStop}%
\bibitem [{\citenamefont {Freese}\ \emph {et~al.}(2013)\citenamefont {Freese},
  \citenamefont {Lisanti},\ and\ \citenamefont {Savage}}]{Freese2013}%
  \BibitemOpen
  \bibfield  {author} {\bibinfo {author} {\bibfnamefont {K.}~\bibnamefont
  {Freese}}, \bibinfo {author} {\bibfnamefont {M.}~\bibnamefont {Lisanti}}, \
  and\ \bibinfo {author} {\bibfnamefont {C.}~\bibnamefont {Savage}},\ }\href
  {\doibase 10.1103/revmodphys.85.1561} {\bibfield  {journal} {\bibinfo
  {journal} {Reviews of Modern Physics}\ }\textbf {\bibinfo {volume} {85}},\
  \bibinfo {pages} {1561} (\bibinfo {year} {2013})}\BibitemShut {NoStop}%
\end{thebibliography}%

\end{document}